\documentclass[11pt,letterpaper]{article}

\usepackage{amsmath,amssymb,amsfonts,amsthm,bbm}

\usepackage{mathtools}   
\usepackage{stmaryrd} 

\usepackage{enumitem}   

\usepackage{authblk} 

\usepackage[titletoc,title]{appendix}  

\usepackage[normalem]{ulem}

\usepackage{kpfonts}
\usepackage[T1]{fontenc}

\usepackage[table,xcdraw,dvipsnames]{xcolor}   
\usepackage{hyperref}
\newcommand\myshade{85}
\colorlet{mylinkcolor}{YellowOrange}
\colorlet{mycitecolor}{Aquamarine}
\colorlet{myurlcolor}{violet}

\hypersetup{
  linkcolor  = mylinkcolor!\myshade!black,
  citecolor  = mycitecolor!\myshade!black,
  urlcolor   = myurlcolor!\myshade!black,
  colorlinks = true,
}

\usepackage{caption}
\usepackage{subcaption}
\usepackage{adjustbox}
\usepackage{multirow}
\usepackage{floatrow}

\usepackage[linesnumbered,ruled,vlined]{algorithm2e}

\SetCommentSty{mycommfont}

\SetKwInput{KwInput}{Input}                
\SetKwInput{KwOutput}{Output}              

\usepackage{listings}             
\usepackage{stmaryrd} 

\renewcommand{\hat}{\widehat}
\renewcommand{\tilde}{\widetilde}
\renewcommand{\bar}{\overline}

\def\bbone{\mathbbm{1}} 

     \def\PP{\mathbb{P}}

\def\calB{{\cal  B}} 
 
 \def\cD{{\cal  D}}
 
 \def\cF{{\cal  F}}

\def\calM{{\cal  M}} \def\cM{{\cal  M}}
 \def\cN{{\cal  N}}

\def\calX{{\cal  X}} 
 
 \def\cZ{{\cal  Z}}

\newcommand{\bfsym}[1]{\ensuremath{\boldsymbol{#1}}}

              \def\bSigma{\bfsym \Sigma}

\providecommand{\norm}[1]{\left\lVert#1\right\rVert}

\providecommand{\angles}[1]{\left\langle #1 \right\rangle}
\providecommand{\paran}[1]{\left( #1 \right)}

\usepackage{mathtools}
\DeclarePairedDelimiterX{\infdivx}[2]{(}{)}{%
  #1 \; \delimsize\| \; #2%
}

\DeclareMathOperator{\Tr}{Tr}
\DeclareMathOperator{\vect}{vec}

\newtheorem{definition}{Definition}[section]
\newtheorem{assumption}[definition]{Assumption}
\newtheorem{lemma}[definition]{Lemma}
\newtheorem{proposition}[definition]{Proposition}
\newtheorem{theorem}[definition]{Theorem}

\newtheorem{example}{Example}

\theoremstyle{definition}
\newtheorem{remark}[definition]{Remark}

\definecolor{royalpurple}{rgb}{0.47, 0.32, 0.66}
\definecolor{greenfresh}{HTML}{00897B}
\definecolor{bluefresh}{HTML}{1E88E5}
\definecolor{redfresh}{HTML}{E53935}

\def\magenta\color{magenta}

\definecolor{royalpurple}{rgb}{0.47, 0.32, 0.66}

\def\beq{\begin{equation}}
\def\eeq{\end{equation}}

\def\bet{\begin{theorem}}
\def\eet{\end{theorem}}

\def\bel{\begin{lemma}}
\def\eel{\end{lemma}}

\usepackage{setspace}
\usepackage{etoolbox}
\BeforeBeginEnvironment{equation*}{\begin{singlespace}\vspace*{-\baselineskip}}
	\AfterEndEnvironment{equation*}{\end{singlespace}\noindent\ignorespaces}
\BeforeBeginEnvironment{equation}{\begin{singlespace}\vspace*{-\baselineskip}}
	\AfterEndEnvironment{equation}{\end{singlespace}\noindent\ignorespaces}
\BeforeBeginEnvironment{align}{\begin{singlespace}\vspace*{-\baselineskip}}
	\AfterEndEnvironment{align}{\end{singlespace}\noindent\ignorespaces}
\BeforeBeginEnvironment{align*}{\begin{singlespace}\vspace*{-\baselineskip}}
	\AfterEndEnvironment{align*}{\end{singlespace}\noindent\ignorespaces}
\BeforeBeginEnvironment{eqnarray}{\begin{singlespace}\vspace*{-\baselineskip}}
	\AfterEndEnvironment{eqnarray}{\end{singlespace}\noindent\ignorespaces}
\BeforeBeginEnvironment{eqnarray*}{\begin{singlespace}\vspace*{-\baselineskip}}
	\AfterEndEnvironment{eqnarray*}{\end{singlespace}\noindent\ignorespaces}

\usepackage[framemethod=TikZ]{mdframed} 

\usepackage{fancybox}

\usepackage[top=1in, bottom=1in, left=1in, right=1in]{geometry}

\usepackage{setspace}
\usepackage[authoryear]{natbib}  
\newcommand{\mybibsty}{chicago}

\usepackage{graphicx}
\graphicspath{{figs/}}

\usepackage{comment}

\DeclareMathOperator*{\matk}{{\text{mat}}_m}

\renewcommand{\hat}{\widehat}
\renewcommand{\tilde}{\widetilde}

\begin{document}
%
%
\newcommand{\TITLE}{Tensor Neyman–Pearson Classification: \\ Theory, Algorithms, and Error Control}

\newcommand{\blind}{0}

\if0\blind
{
	\title{\bf \TITLE}
    \author{
    Lingchong~Liu$^\sharp$ \hspace{8ex}
    Elynn~Chen$^\dag$ \hspace{8ex}
    Yuefeng~Han$^\flat$ \hspace{8ex}
    Lucy~Xia$^\sharp$\thanks{L.~Xia is the corresponding author: lucyxia@ust.hk. Co-advisors are ordered alphabetically. E.~Chen is supported in part by NSF Grants DMS-2412577; Y.~Han is supported in part by Grants DMS-2412578.}  \\ \normalsize
    $^{\sharp}$The Hong Kong University of Science and Technology  \\
    $^\dag$New York University \hspace{8ex}
    $^\flat$ University of Notre Dame \hspace{8ex}
    }
	\date{\today}
	\maketitle
} \fi

\if1\blind
{
	\bigskip
	\bigskip
	\bigskip
	\title{\bf ...}
	\date{\vspace{-5ex}}
	\maketitle
	\medskip
} \fi


%
%
\begin{abstract}
Biochemical discovery increasingly relies on classifying molecular structures when the consequences of different errors are highly asymmetric. In mutagenicity and carcinogenicity, misclassifying a harmful compound as benign can trigger substantial scientific, regulatory, and health risks, whereas false alarms primarily increase laboratory workload. Modern representations transform molecular graphs into persistence image tensors that preserve multiscale geometric and topological structure, yet existing tensor classifiers and deep tensor neural networks provide no finite-sample guarantees on type I error and often exhibit severe error inflation in practice.

We develop the first Tensor Neyman-Pearson (Tensor-NP) classification framework that achieves finite-sample control of type I error while exploiting the multi-mode structure of tensor data. Under a tensor-normal mixture model, we derive the oracle NP discriminant, characterize its Tucker low-rank manifold geometry, and establish tensor-specific margin and conditional detection conditions enabling high-probability bounds on excess type II error. We further propose a Discriminant Tensor Iterative Projection estimator and a Tensor-NP Neural Classifier combining deep learning with Tensor-NP umbrella calibration, yielding the first distribution-free NP-valid methods for multiway data. 
Across four biochemical datasets, Tensor-NP classifiers maintain type I errors at prespecified levels while delivering competitive type II error performance, providing reliable tools for asymmetric-risk decisions with complex molecular tensors.
\end{abstract}

\section{Introduction}  \label{sec:intro}

Advances in computational chemistry and molecular biology are producing increasingly rich structural data where scientific decisions hinge on asymmetric classification risks \citep{Morris+2020, nils2012subgraph}. In mutagenicity screening, carcinogenicity prediction, or enzyme inhibitor discovery, a single misclassified toxic or carcinogenic compound may lead to severe health risks, regulatory failures, and costly downstream experiments, whereas a false alarm primarily increases laboratory workload. These applications motivate tensor classifiers that explicitly control the probability of misclassifying harmful compounds as safe or active compounds as inactive while preserving high power.

The molecular datasets we analyze--chemical mutagenicity (MUTAG), cyclooxygenase-2 inhibitor (COX2), benzodiazepine receptor binding (BZR), and chemical carcinogenicity (PTC\_MM)--illustrate the complexity of this task. Each molecule is represented as a chemical graph encoding atom types, bonding patterns, 3-D arrangements, and functional substructures \citep{debnath1991structure, sutherland2003spline, helma2001predictive, Morris+2020}. Transforming these irregular graphs into persistence images yields high-dimensional multiway arrays that capture local geometry, global connectivity, and multiscale topological features \citep{wu2025tensor, hou2024tensor, omberg2009global, yahyanejad2019survey}. These tensor representations are scientifically powerful but amplify the statistical challenge: the covariates are multi-modal, highly structured, and high-dimensional, and the classification task requires stringent control of asymmetric errors.

Existing tensor classifiers, including tensor linear discriminant analysis (LDA), tensor logistic regression, support tensor machines, Tucker/CP discriminant models, and modern tensor neural networks \citep{pan2019covariate, chen2024higha, chen2024highb, wen2024tensor}, exploit multiscale structure to varying degrees, but none provides finite-sample guarantees on type I error. Deep tensor architectures can approximate flexible nonlinear decision boundaries, yet their score functions are highly unstable under class imbalance and small sample sizes. As our empirical analyses show, state-of-the-art tensor neural networks may violate desired type I error levels in 70-90\% of repeated experiments, even when overall accuracy is high. These findings point to a fundamental methodological gap: current approaches either preserve tensor structure without controlling asymmetric errors, or deliver flexible modeling at the expense of statistical reliability.

This paper fills the gap by developing a unified framework for Tensor Neyman-Pearson (Tensor-NP) classification, which provides structure-aware modeling together with finite-sample type I error guarantees. Extending the NP paradigm to tensor-valued predictors requires addressing two major challenges. First, multi-mode covariance structure and tensor covariates fundamentally change the geometry of the oracle NP discriminant: under a tensor-normal mixture model with Kronecker-structured covariance matrices, the optimal NP rule involves a Tucker low-rank discriminant tensor with mode-wise precision operators that vectorization fails to exploit. Second, classical NP arguments based on margin and detection conditions demand new derivations tailored to multiway settings. The tensor structure alters the functional form of score deviations, and induces nontrivial dependencies across modes.

We address these challenges through three contributions. First, we derive the first NP oracle rule for tensor data, establishing its analytic form, and the tensor-specific algebra governing its score tensor geometry. Second, we prove tensor-specific conditional margin and  detection conditions, constructing a high-probability region that simultaneously controls score deviations for both classes and generalizing prior NP theory beyond vector settings. 
Third, we design structure-preserving estimators and algorithms that attain the theoretical guarantees: in the parametric regime, the Discriminant Tensor Iterative Projection (DTIP) estimator preserves Tucker low-rank manifold geometry and achieves finite-sample NP oracle inequalities with rates adapting to Tucker ranks and mode dimensions; in the nonparametric regime, the Tensor Neural NP Classifier integrates tensor contraction layers with NP umbrella calibration to obtain the first distribution-free, model-agnostic, NP classifier for multiway data.

These contributions clarify how this work advances both statistics and artificial intelligence (AI). From statistics to AI, we show that classical decision theory can discipline modern deep learning models by endowing their outputs with provable, high-probability guarantees: NP calibration stabilizes the tail behavior of neural scores, and tensor-specific margin and detection conditions reveal how low-rank tensor manifold geometry influences the representation of score functions. From AI to statistics, tensor neural networks enrich statistical methodology by providing accurate nonlinear scoring functions in regimes where linear tensor discriminants are insufficient, thereby enabling NP classifications in structurally complex, high-dimensional domains. The tensor contraction layers, in particular, create low-dimensional yet structure-preserving representations that integrate seamlessly into the NP calibration pipeline.

Together, these developments produce the first practically deployable framework for Neyman-Pearson classification on tensor-valued data. By uniting tensor discriminant analysis, deep tensor architectures, and rigorous asymmetric-risk control, our approach provides statistically reliable and scientifically interpretable tools for modern biochemical applications and, more broadly, for high-stakes decisions involving complex multiway data.


\subsection{Related work and our distinction} \label{sec:relatedwork}

\noindent
\textbf{Neyman--Pearson classification foundations.}
Research under the NP classification paradigm \citep{scott2005neyman, rigollet2011neyman} can be broadly categorized into three lines. 
(i) \emph{Foundational theory:} this includes the convex formulation of the NP problem and oracle inequalities \citep{rigollet2011neyman}, performance characterizations via NP--ROC bands \citep{tong2018neyman}, and recent developments on distribution-free and minimax optimal rates \citep{kalan2024distribution}.  
(ii) \emph{Methodological advances:} plug-in classifiers based on density-ratio estimation \citep{tong2013plug, zhao2016neyman}, model-based NP classifiers including linear discriminant analysis \citep{tong2020neyman, wang2024non}, and sample-splitting based umbrella algorithms \citep{tong2018neyman}.  
(iii) \emph{Extensions to structured or complex settings:} domain adaptation \citep{scott2019generalized}, hierarchical and multi-stage decisions \citep{wang2024hierarchical}, fairness and cost-sensitive learning \citep{fan2023neyman, tian2024neyman}, transfer learning \citep{kalan2025transfer}, and handling noisy labels \citep{yao2023asymmetric}.  
Comprehensive overviews are provided in recent surveys such as \citet{tong2016survey}.  
Despite these advances, all existing NP methods operate on \emph{vector inputs} and rely on scalar or linear-score models; the extension of NP guarantees to multiway structured data remains unexplored.

\smallskip
\noindent
\textbf{Tensor classification and regression.}
Tensor learning has become a core methodology for high-dimensional structured data \citep{zhang2018tensor,chen2020semiparametric,chen2025distributed}.  
In regression, most works impose Tucker low-rankness to achieve parsimonious estimation and interpretability \citep{HanWillettZhang2022,xu2025statistical}, while CP decompositions are comparatively underdeveloped beyond the early block relaxation approach of \citet{zhou2013tensor}.  
In classification, representative models include support tensor machines \citep{hao2013linear}, tensor logistic regression \citep{zhou2013tensor}, tensor discriminant analysis and its variants \citep{pan2019covariate,chen2024higha,chen2024highb}, as well as nonlinear approaches such as tensor neural networks and kernelized discriminant analysis \citep{wen2024tensor}.
Recent progress has extended tensor discriminant models to high-dimensional settings with theoretical guarantees \citep{chen2024higha}, and to covariate-adjusted scenarios through the CATCH framework \citep{pan2019covariate}.  
Nevertheless, all these methods aim to minimize overall misclassification risk and do not provide user-specified control of asymmetric errors.


\smallskip
\noindent
\textbf{Challenges and our distinction.} In summary, the challenges in extending the Neyman--Pearson paradigm to tensor-valued data arise from the intrinsic low-rank manifold and complexity of multi-mode structures. Unlike vectors, tensors induce coupled covariance dependencies across modes, making classical reductions through vectorization inefficient and forcing a rethinking of the oracle rule itself. Verifying margin and detection conditions also becomes conceptually harder and different. On the methodological side, preserving low-rank tensor structure while maintaining type I error control demands estimators and algorithms that respect both statistical efficiency and low-rank manifold constraints, something existing NP classifiers were not designed to handle. These distinctions underscore why the tensor setting is not a straightforward extension of prior work, but instead requires new theory, conditions, and procedures to reconcile asymmetric-risk control with tensor-aware modeling.

\subsection{Notation and Organization}

Vectors, matrices, and tensors are denoted by lowercase letters ($x$), uppercase letters ($X$), and calligraphic letters ($\mathcal{X}$), respectively.  
Let $[N]=\{1,\ldots,N\}$ and $\bbone(\cdot)$ denote the indicator function.  
For sequences $\{a_n\}$ and $\{b_n\}$, write $a_n=O(b_n)$ if $|a_n|\le C|b_n|$, $a_n\asymp b_n$ if $C^{-1}\le a_n/b_n\le C$, $a_n \lesssim b_n$ if $a_n\le C b_n$, and $a_n=o(b_n)$ if $a_n/b_n\to0$, for some constant $C>0$.  

For a matrix $X$, let $X_{i\cdot}$, $X_{\cdot j}$, and $X_{ij}$ denote its $i$-th row, $j$-th column, and $(i,j)$-th entry.  
We write $\lambda_{\min}(X)$ and $\lambda_{\max}(X)$ for the smallest and largest eigenvalues, $\sigma_i(X)$ for the $i$-th largest singular value, $\|X\|_F$ and $\|X\|$ for the Frobenius and spectral norms, and $\Tr(X)$ for the trace.  
For a tensor $\mathcal{X}$, $\vect(\mathcal{X})$ denotes vectorization and $\mathrm{mat}_k(\mathcal{X})$ denotes its mode-$k$ unfolding.  
The inner product is $\langle\mathcal{X},\mathcal{Y}\rangle = \vect(\mathcal{X})^\top\vect(\mathcal{Y})$, inducing the Frobenius norm $\|\mathcal{X}\|_F = \langle\mathcal{X},\mathcal{X}\rangle^{1/2}$.  
The mode-$k$ product $\mathcal{X}\times_k A$ between $\mathcal{X}\in\mathbb{R}^{d_1\times\cdots\times d_M}$ and $A\in\mathbb{R}^{\tilde d_k\times d_k}$ produces a tensor of size $d_1\times\cdots\times d_{k-1}\times\tilde d_k\times d_{k+1}\times\cdots\times d_M$ with entries  
$(\mathcal{X}\times_k A)_{i_1\cdots i_{k-1} j i_{k+1}\cdots i_M} = \sum_{i_k=1}^{d_k}\mathcal{X}_{i_1\cdots i_k\cdots i_M}A_{j i_k}$.  
We denote $d=\prod_{k=1}^M d_k$ and $d_{-k}=d/d_k$.  
See \citet{kolda2009tensor} for further tensor algebraic notation.

An $M$-th order tensor $\mathcal{X}\in\mathbb{R}^{d_1\times\cdots\times d_M}$ follows a tensor-normal distribution $\mathcal{X}\sim\mathcal{TN}(\mathcal{M};\bSigma)$ with mean tensor $\mathcal{M}$ and mode-specific covariance matrices $\bSigma=(\Sigma_1,\ldots,\Sigma_M)$, where each $\Sigma_m$ is $d_m\times d_m$, if and only if $\vect(\mathcal{X})\sim\mathcal{N}(\vect(\mathcal{M});\Sigma_M\otimes\cdots\otimes\Sigma_1)$ and $\otimes$ denotes the Kronecker product.

\smallskip
\noindent
\textbf{Organization.}
Section~\ref{sec:framework} introduces the tensor NP framework and algorithms for both parametric and nonparametric tensor classification.  
Section~\ref{sec:theory} presents the theoretical analysis for the parametric setting, including preliminary lemmas, the conditional margin and detection conditions, and oracle inequalities.  
Sections~\ref{sec:simulation} and~\ref{sec:application} report simulation studies and application to three molecular mutagenicity datasets, respectively.  
Section~\ref{sec:conclusion} concludes.  
All technical proofs, additional numerical results on simulations and an extra molecular dataset are provided in Supplementary Material.

\section{Tensor Neyman--Pearson Framework} \label{sec:framework}
This section formalizes tensor Neyman--Pearson (NP) classification and establishes that constructing a sample-based NP classifier reduces to two key components: 
(i) $\widehat{s}(\cdot)$, an estimator of the scoring function $s(\cdot)$ that ranks observations by their evidence for class~1, and 
(ii) $\widehat{C}_\alpha$, an estimator of a threshold $C_\alpha$, that ensures with high probability, the population type I error rate does not exceed a pre-specified level~$\alpha$. 
We instantiate $\widehat{s}(\cdot)$ using both parametric and nonparametric tensor-based discriminant models, and determine $\hat{C}_\alpha$ through the NP umbrella algorithm to ensure valid finite-sample error control with high probability.

\subsection{Problem Setup and NP Formulation} \label{sec:setup}

Let $(\mathcal{X},Y)$ be a pair where $\mathcal{X}\in\mathbb{R}^{d_1\times\cdots\times d_M}$ is an $M$-th order tensor and $Y\in\{0,1\}$ is a class label. For a measurable classifier $\phi:\mathbb{R}^{d_1\times\cdots\times d_M}\to\{0,1\}$, define type I and type II errors by $R_0(\phi)=\Pr(\phi(\mathcal{X})=1\mid Y=0)$ and $R_1(\phi)=\Pr(\phi(\mathcal{X})=0\mid Y=1)$, respectively. 
The NP paradigm seeks the classifier that minimizes the type II error while controlling the type I error at a prescribed level $\alpha\in(0,1)$:
\begin{equation} \label{eq:np_opt} 
\min_{\phi}~R_1(\phi)\quad\text{subject to}\quad R_0(\phi)\le\alpha,\quad \alpha\in(0,1).
\end{equation}
By the Neyman--Pearson lemma, the NP oracle classifier can be expressed through the density ratio as
\begin{equation*}
\phi^*_\alpha(\mathcal{X})=\bbone\!\left\{\frac{f_1(\mathcal{X})}{f_0(\mathcal{X})}>C^*_\alpha\right\},
\end{equation*}
where $f_1(\mathcal{X})$ and $f_0(\mathcal{X})$ denote the conditional densities of $\mathcal{X}$ given $Y=1$ and $Y=0$, respectively.  The oracle threshold $C^*_\alpha$ is determined by
\begin{equation*}
\Pr\!\left(\tfrac{f_1(\mathcal{X})}{f_0(\mathcal{X})}>C^*_\alpha\mid Y=0\right)\le\alpha,
\qquad
\Pr\!\left(\tfrac{f_1(\mathcal{X})}{f_0(\mathcal{X})}\ge C^*_\alpha\mid Y=0\right)\ge\alpha.
\end{equation*}
To accommodate a broad class of scoring-based classifiers, we define a general population \emph{scoring function} $s^*(\mathcal{X})$, which is strictly increasing in $f_1(\mathcal{X})/f_0(\mathcal{X})$. The corresponding \emph{NP oracle classifier} is obtained as follows (with slight abuse of double using the notation $C^*_\alpha$):
\begin{equation} \label{eq:np_rule}\footnotesize
\begin{aligned}
\phi^*_\alpha(\mathcal{X})=\bbone\{\,s^*(\mathcal{X})>C^*_\alpha\,\}, \ \ \text{with}\ \ \Pr(s^*(\mathcal{X})>C^*_\alpha\mid Y=0)\leq \alpha,\quad \Pr\!\left(s^*(\mathcal{X})\ge C^*_\alpha\mid Y=0\right)\ge\alpha.
\end{aligned}
\end{equation}

In practice, the distributions $f_0$ and $f_1$ are unknown, and constructing an NP classifier from data therefore requires estimating both the scoring function $s^*(\cdot)$ and a threshold $C^*_\alpha$.  
The resulting sample-level classifier
$
\hat{\phi}_\alpha(\mathcal{X})=\bbone\{\hat{s}(\mathcal{X})>\hat{C}_\alpha\}
$
should control the type I error at level~$\alpha$ with high probability $1-\delta$ while achieving a small excess type II error. We refer to \(\delta\) as the violation rate. It is worth noting that adopting high-probability type I error control is more appropriate in asymmetric error settings than controlling the type I error only in expectation. Under expectation-based control, there remains a substantial probability (e.g., around 50\%) that the realized type I error exceeds the target level, which can incur substantial costs in practice. The next subsections develop these two components: Section~\ref{sec:score_models} constructs $\hat{s}(\cdot)$ under tensor-based discriminant models, and Section~\ref{sec:umbrella} calibrates $\hat{C}_\alpha$ via the NP umbrella algorithm adapted especially for tensor observations.

\subsection{Tensor-Based Scoring Functions} \label{sec:score_models}

We estimate the scoring function $s(\cdot)$ using two complementary classifiers that preserve the multi-mode tensor structure: a parametric \emph{tensor linear discriminant analysis} (T-LDA) model, which enables a thorough theoretical characterization and excels in the small-sample regime, and a nonparametric, universal \emph{tensor neural network} (T-NN) model, which substantially broadens the applicability of our framework by accommodating nonlinear decision boundaries and high-order interactions. Both aim to approximate the oracle discriminant while exploiting the intrinsic low-rank geometry of tensor.

\subsubsection{Tensor LDA-NP Classifier} \label{sec:T-LDA-NP}

In the parametric regime, we assume that $\mathcal{X}\mid Y=y$ follows a tensor-normal distribution with class-specific means and a common Kronecker-structured covariance.  
Although seemingly restrictive, vector LDA and its extensions perform well even in high-dimensional and mildly misspecified settings \citep{chen2024higha}; their tensor counterparts inherit the same interpretability and analytical tractability.  
The tensor LDA model provides an explicit parameterization that yields a closed-form oracle rule, making it particularly suitable for theoretical analysis. Formally, we specify
\begin{equation} \label{eqn:tgmm2}
   \mathcal{X}\mid Y=y \sim \mathcal{TN}(\mathcal{M}_y;\bSigma), \qquad
   \pi_y=\Pr(Y=y), \quad \pi_0+\pi_1=1, \quad y\in\{0,1\},
\end{equation}
where $\bSigma=(\Sigma_1,\ldots,\Sigma_M)$ denotes the mode-specific covariance matrices.  
Under this model, the NP oracle classifier can be written as\footnote{Throughout, $C_\alpha$ denotes the oracle threshold for a general scoring function, $C_\alpha^\ast$ the threshold for the density-ratio score, and $C_\alpha^{\ast\ast}$ that for the linear score in T-LDA.}
\begin{equation}\label{eqn:oracle classifier}
\phi_\alpha^\ast(\mathcal{X})
= \bbone\!\left\{\frac{f_1(\mathcal{X})}{f_0(\mathcal{X})}>C_\alpha^\ast\right\}
= \bbone\!\left\{\exp\!\big(\langle \mathcal{X}-\mathcal{M},\mathcal{B}\rangle\big)>C_\alpha^\ast\right\}
= \bbone\!\left\{\langle \mathcal{X},\mathcal{B}\rangle>C_\alpha^{\ast\ast}\right\},
\end{equation}
where $\cM = (\cM_0 + \cM_1)/2$, discriminant tensor $\calB = \cD \times_{m=1}^M \Sigma_m^{-1}$ with $\cD=\cM_1-\cM_0$, and $C_\alpha^{\ast\ast}=\log C_\alpha^\ast + \langle \cM, \; \calB \rangle$ is a constant threshold depending on $\alpha$. The signal-to-noise ratio (SNR) can be represented by $\Delta=\sqrt{\langle \calB, \; \cD \rangle}$.

\begin{remark}\label{rmk:classical-optimal-classifier}
If the objective is to minimize the overall classification error rather than controlling the type I error, by Bayes' theorem, we have $C_\alpha^{\ast\ast} = \langle \mathcal{M}, \mathcal{B} \rangle - \log(\pi_1/\pi_0)$, which does not depend on $\alpha$. In this case, \eqref{eqn:oracle classifier} reduces to Fisher’s LDA rule when $M=1$.
\end{remark}

Let $s^\ast(\mathcal{X})=\langle\mathcal{X},\mathcal{B}\rangle$ denote the oracle score.  
Its distribution and the corresponding threshold $C_\alpha^{\ast\ast}$ are summarized below.

\begin{lemma}\label{lem:oracle score function}
Under the tensor LDA model \eqref{eqn:tgmm2} and the oracle classifier \eqref{eqn:oracle classifier}, the oracle score $s^\ast(\mathcal{X})$ and threshold $C_\alpha^{\ast\ast}$ satisfy
\begin{equation}\label{eqn:oracle score function}
\begin{aligned}
s^\ast(\mathcal{X}) &\sim \mathcal{N}\!\left(\vect(\mathcal{B})^\top\vect(\mathcal{M}_0),~
\vect(\mathcal{D})^\top\bSigma_v^{-1}\vect(\mathcal{D})\right),
\quad \text{for } \mathcal{X}\sim\text{class }0,\\
C_\alpha^{\ast\ast} &= \sqrt{\vect(\mathcal{D})^\top\bSigma_v^{-1}\vect(\mathcal{D})}\,
\Phi^{-1}(1-\alpha) + \vect(\mathcal{B})^\top\vect(\mathcal{M}_0),
\end{aligned}
\end{equation}
where $\Phi(\cdot)$ denotes the standard normal CDF and $\bSigma_v=\Sigma_M\otimes\cdots\otimes\Sigma_1$.
\end{lemma}
The NP oracle classifier \eqref{eqn:oracle classifier} exploits the low-rank structure by employing a discriminant tensor 
$\mathcal{B}$ with a Tucker decomposition:
\begin{equation}\label{eqn:lda-tucker}
\mathcal{B}=\mathcal{F}\times_1 U_1\times_2\cdots\times_M U_M,
\end{equation}
with Tucker rank $(r_1,\ldots,r_M)$, $\mathcal{F}$ as an $r_1\times\cdots\times r_M$ core tensor, and each $U_m\in\mathbb{R}^{d_m\times r_m}$ is orthonormal.  To estimate $\mathcal{B}$, we develop in Algorithm \ref{alg:tensorlda-tucker}, the Discriminant Tensor Iterative Projection (DTIP) method. Starting with a spectral initialization which guarantees the Tucker low-rankness \citep{zhang2018tensor}, we apply iterative orthogonal projection refinements to achieve more accurate low-rank tensor estimates. Intuitively, at each iteration, projecting the tensor onto the orthonormal spaces of all other modes effectively denoises the signal, allowing the algorithm to update the orthonormal matrix for the current mode. More specifically, the spectral initialization is given by
\begin{equation}\label{eqn:lda-discrim-tensor}
\hat{\mathcal{B}}^{\mathrm{init}}=\big(\hat{\cM_1}-\hat{\cM_0}\big)\times_{m=1}^M\hat{\Sigma}_m^{-1},
\end{equation}
where $n_y$ is the sample size of class $y$, 
$\hat{\cM_y}=(n_y)^{-1}\sum_{i=1}^{n_y}\mathcal{X}_i^{(y)}$ is the sample mean, and
\begin{equation*}
\hat\Sigma_m
= (n d_{-m})^{-1}\sum\nolimits_{y=0}^1\sum\nolimits_{i=1}^{n_{y}}\matk\left(\calX_{i}^{(y)}-\hat{\cM_y}\right){\matk}^{\top}\left(\calX_{i}^{(y)}-\hat{\cM_y}\right), \quad\text{for } m\in[M],
\end{equation*}
with $\matk(\calX_i^{(y)})$ denoting the mode-$m$ unfolding of $\mathcal{X}_i^{(y)}$ into a $d_m\times d_{-m}$ matrix with $d=\prod_{m=1}^M d_m, d_{-m}=d/d_m$ and $n=n_{0}+n_{1}$. 
Class probabilities are estimated as $\hat{\pi}_y=n_y/(n_0+n_1)$, for $y=0,1$.

The final estimator $\hat{\calB}$ is obtained via orthogonal projection refinement.  
Plugging it into the classical discriminant rule yields the estimated classifier T-LDA, without addressing the control over the type I error,
\begin{equation}\label{eqn:lda-rule-tucker}
\hat{\phi}^{\text{ LDA}}(\mathcal{X})
=\bbone\!\left\{
\langle \mathcal{X}-(\hat{\cM_0}+\hat{\cM_1})/2,\;
\hat{\calB}\rangle
+\log(\hat{\pi}_1/\hat{\pi}_0)\ge0
\right\},
\end{equation}
where we treat $\hat{s}^{\text{ LDA}}(\mathcal{X})=\langle\mathcal{X},\hat{\calB}\rangle$ as the estimated score function and $\hat{C}^{\text{ LDA}} =  \langle (\hat{\cM_0}+\hat{\cM_1})/2,\;
\hat{\calB}\rangle
-\log(\hat{\pi}_1/\hat{\pi}_0)$ as the threshold.

We further construct the T-LDA-NP classifier (Algorithm~\ref{alg:T-LDA-NP}) 
\[
\hat{\phi}_\alpha^{\text{ LDA-NP}}(\mathcal{X}) = \bbone\{\hat{s}^{\text{ LDA-NP}}(\mathcal{X})> \hat{C}_\alpha^{\text{ LDA-NP}},\}
\]
by updating the scoring function and the threshold, where the only difference between $\hat{s}^{\text{ LDA-NP}}$ and $\hat{s}^{\text{ LDA}}$ is that half of the class 0 training data are held out for threshold calibration and construction of $\hat{C}_\alpha^{\text{ LDA-NP}}$. Please refer to  Section~\ref{sec:umbrella} for more details on $\hat{C}_\alpha^{\text{ LDA-NP}}$. 

\begin{algorithm}[htpb!]
\SetKwInOut{Input}{Input}
\SetKwInOut{Output}{Output}
\Input{Initial tensor $\hat{\calB}^{\text{init}}$, 
Tucker rank $(r_1,...,r_M)$, tolerance parameter $\epsilon>0$, maximum  iterations $T$}
\Output{$\hat{\calB}=\hat{\calB}^{\text{init}}\times_{m=1}^M \hat  U_{m}\hat  U_{m}^\top$, and 
$\hat \cF = \big|\hat{\calB}^{\text{init}}\times_{m=1}^M (\hat  U_{m})^\top\big|$,
where $\hat  U_{m}=\hat U_{m}^{(t)}$.}

Let $t=0$, initiate via high order SVDs, $\hat U_m^{(0)}={\rm LSVD}_{r_m}({\rm mat}_m(\hat{\calB}^{\text{init}})),\ 1\le m\le M$, where LSVD$_{r_m}$ represents top $r_m$ left singular vectors

\Repeat{$t = T$ {\bf or} $\max_{m}\|\hat U_{m}^{(t)}(\hat U_{m}^{(t)})^\top- \hat U_{m}^{(t-1)}(\hat U_{m}^{(t-1)})^\top\|_{2}\le \epsilon$}{
Set $t=t+1$. 

\For{$m = 1$ to $M$}{
Calculate $\cZ_m=\hat{\calB}^{\text{init}} \times_1 \hat U_{1}^{(t)\top} \times_2 \cdots \times_{m-1} \hat U_{m-1}^{(t)\top} \times_{m+1} \hat U_{m+1}^{(t-1)\top} \times_{m+2}\cdots\times_M \hat U_{M}^{(t-1)\top}$

Perform high order SVD,
$\hat U_m^{(t)}={\rm LSVD}_{r_m}({\rm mat}_m(\cZ_m))$.
}
}
\caption{Discriminant Tensor Iterative Projection (DTIP) with Tucker Low-Rank}
\label{alg:tensorlda-tucker}
\end{algorithm}

\subsubsection{Tensor Neural Network NP Classifier (T-NN-NP)} \label{sec:T-NN}

To capture nonlinear decision boundaries while preserving multi-mode tensor structure, we construct a \emph{Tensor Neural Network NP Classifier (T-NN-NP)}. 
The model learns a scoring function $\hat s^{\text{ NN}}(\mathcal{X})$ parameterized by a neural network with tensor contraction layers (TCLs) that reduce the input tensor to a compact low-rank core before standard fully connected layers \citep{wen2024tensor}.  
Let $h_\theta(\mathcal{X})\in[0,1]$ denote the network's predicted probability of class~1; we define the scoring function as $\hat s^{\text{ NN}}(\mathcal{X})=g(h_\theta(\mathcal{X}))$, where $g$ is a monotone link (e.g., identity or logit).  
This design ensures that the NP threshold calibration in Section~\ref{sec:umbrella} applies directly to $\hat s^{\text{ NN}}(\cdot)$, yielding valid finite-sample type I control for a broad class of neural architectures.

Unlike conventional deep neural networks that flatten tensors into vectors, TCLs perform mode-wise projections that preserve multiway dependencies and reduce parameters dramatically.  
Given an activation tensor $\mathcal{X}\in\mathbb{R}^{N\times D_1\times\cdots\times D_M}$, where $N$ is the mini-batch size and $D_m$ is the size of mode $m$, each TCL produces a compact tensor
$
\mathcal{X}'=\mathcal{X}\times_2 V_1\times_3 V_2\times\cdots\times_{M+1} V_M,
$
where $V_m\in\mathbb{R}^{R_m\times D_m}$ are learnable projection matrices with $R_m\ll D_m$.  
The contraction begins at mode~2, preserving the batch dimension (mode~1).  
These projection matrices are optimized jointly with the network parameters via backpropagation, allowing the TCLs to discover the most informative low-dimensional tensor representations for classification \citep{kossaifi2020tensor,wen2024tensor}.

\begin{figure}[ht]
    \centering
    \includegraphics[width=0.9\linewidth]{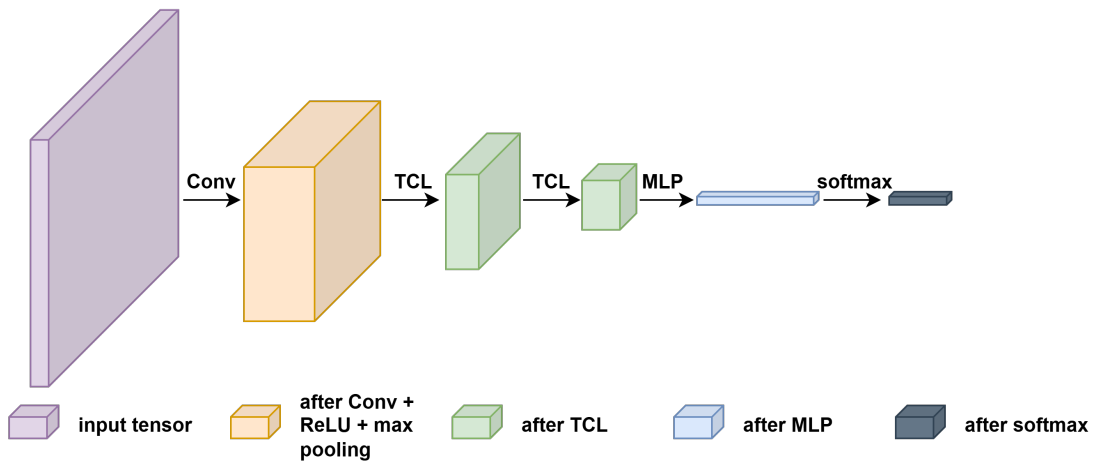}
    \caption{ The architecture diagram of the tensor neural network model in T-NN and T-NN-NP}
    \label{fig:T-NN}
\end{figure}

The complete architecture, illustrated in FIG~\ref{fig:T-NN}, integrates TCLs with a multilayer perceptron (MLP) classifier, whose output serves as a valid scoring function for NP calibration. 
In this way, T-NN-NP unites the expressive power of deep tensor networks with the statistical rigor of the NP framework: 
the neural network flexibly learns classifier $\hat{s}^{\text{ NN}}$ , while the tensor-adapted NP umbrella algorithm guarantees type I control at level~$\alpha$ and 
Algorithm~\ref{alg:T-NN-NP} summarizes the training procedure.  In short, the T-NN classifier is defined as $\bbone\{\hat{s}^{\text{ NN}}(\calX) > 0.5\}$, while the T-NN-NP classifier is defined as $\bbone\{\hat{s}^{\text{ NN}}(\calX) > \hat{C}_\alpha^\text{ NN-NP}\}$, where $\hat{C}_\alpha^\text{ NN-NP}$ is determined by the tensor-adapted NP umbrella algorithm in Section~\ref{sec:umbrella}.

\begin{remark}[Statistical interpretation]\label{rmk:tnn_np}
The T-NN-NP classifier can be viewed as a nonparametric plug-in NP classification procedure.  
The tensor-adapted NP umbrella algorithm converts the estimated neural network component $\hat{s}^{\text{ NN}}(\mathcal{X})$ into a decision rule that satisfies 
$\Pr(\hat{s}^{\text{ NN}}(\mathcal{X})>\hat{C}_\alpha\mid Y=0)\le\alpha$ with high probability.  
Hence, T-NN-NP inherits both the expressive power of deep low-rank tensor architectures and the distribution-free type I control guaranteed by the NP umbrella algorithm, forming a general plug-in template for nonlinear asymmetric-risk classification.
\end{remark}

\subsection{Tensor-adapted Threshold Calibration via the Umbrella Algorithm} \label{sec:umbrella}

In the newly proposed tensor NP classifiers, including T-LDA-NP and T-NN-NP, the final step is to determine a threshold that enforces the desired type I error bound while accounting for the high-dimensional, multi-mode structure of the data.  
Although the oracle threshold \(C_{\alpha}^{\ast\ast}\) in the tensor LDA model admits a closed-form expression, its plug-in estimator cannot guarantee finite-sample, high-probability control of the type~I error. The key difficulty is that, no matter how accurately \(C_{\alpha}^{\ast\ast}\) is estimated, the estimation error is symmetric; consequently, around half of the time, the estimator may underestimate the true threshold. Such underestimation directly inflates the rejection region, making it impossible to ensure that the type~I error is controlled at the desired level with high probability. For tensor neural network models, where $\hat{s}(\mathcal{X})$ is a highly nonlinear function of multiway inputs, the score distribution is analytically intractable.  
Hence, the calibration step must be \emph{model-free yet structure-aware}, ensuring valid type I error control regardless of how the score is obtained.

We adapt the \emph{NP umbrella algorithm} \citep{tong2018neyman} to the tensor setting as a unified calibration layer that decouples score estimation from statistical error control.  
The algorithm takes any tensor-based score $\hat{s}(\mathcal{X})$, parametric or nonparametric, and determines a data-driven threshold $\hat{C}_\alpha$ that guarantees $\Pr(R_0(\hat{\phi}_\alpha)>\alpha)\le\delta$.  
To accommodate the dependence structure in tensor-valued covariates, calibration is performed on an independent holdout subset of the class 0 samples, ensuring that cross-mode correlations estimated in training do not bias threshold selection. As summarized in Algorithm~\ref{alg:umbrella} in Supplementary Material, we partition the class 0 data into two disjoint subsets $\mathcal{S}_0'$ and $\mathcal{S}_0''$ (for instance, of equal sizes), while all class 1 samples form $\mathcal{S}_1$.  
The score $\hat{s}(\cdot)$ obtained from either T-LDA-NP or T-NN-NP is fitted on $\mathcal{S}_0'\cup\mathcal{S}_1$ and evaluated on $\mathcal{S}_0''$ to produce calibration scores $T=\{\hat{s}(\mathcal{X}):\mathcal{X}\in\mathcal{S}_0''\}$.  
Let $t_{(1)}\le\cdots\le t_{(n_0'')}$ denote the ordered values in $T$ with $n_0''=|\mathcal{S}_0''|$.  
For a violation rate $\delta\in(0,1)$, define
\begin{equation}\label{eq:umbrella_k}
k^*=\min\Bigg\{k\in[n_0'']:
\sum_{j=k}^{n_0''}\binom{n_0''}{j}\alpha^{\,j}(1-\alpha)^{n_0''-j}\le\delta
\Bigg\},
\qquad \text{and}\qquad 
\hat{C}_\alpha=t_{(k^*)},
\end{equation}
and construct the classifier
\begin{equation}\label{eq:umbrella_rule}
\hat{\phi}_\alpha(\mathcal{X})=\bbone\{\hat{s}(\mathcal{X})>\hat{C}_\alpha\}.
\end{equation}
This rule satisfies $\Pr(R_0(\hat{\phi}_\alpha)>\alpha)\le\delta$ under mild continuity assumptions on $\hat{s}(\cdot)$.

\begin{proposition}[\citet{tong2018neyman}]
\label{prop:umbrella}
Let $\hat{s}(\cdot)$ be any real-valued scoring function trained on $\mathcal{S}_0'\cup\mathcal{S}_1$, and let $t_{(1)},\ldots,t_{(n_0'')}$ denote its ordered scores on $\mathcal{S}_0''$.  
Then for the classifier $\hat{\phi}_{k}(X)=\bbone\{\hat{s}(X)>t_{(k)}\}$, the population type I error satisfies
\[
\Pr\{R_0(\hat{\phi}_{k})>\alpha\}\le
\sum_{j=k}^{n_0''}\binom{n_0''}{j}\alpha^{\,j}(1-\alpha)^{n_0''-j}.
\]
\end{proposition}

 With Proposition~\ref{prop:umbrella}, the choice of $k^*$ in \eqref{eq:umbrella_k} ensures that the resulting classifier $\hat{\phi}_\alpha$ in \eqref{eq:umbrella_rule} is able to control the type I error below level $\alpha$ with probability at least $1-\delta$, for any tensor-based score, including those derived from Tucker-structured discriminants and deep tensor neural networks.

\begin{algorithm}[htpb!]
    \caption{T-LDA and T-LDA-NP Classifier}
    \label{alg:T-LDA-NP}
    \SetKwInOut{Input}{Input}
    \SetKwInOut{Output}{Output}
    \SetKwFunction{FMain}{T-LDA}
    \SetKwProg{Fn}{Function}{:}{}
    
    \Input{
      \begin{itemize}
      \item[-]	Randomly split class $0$ training data as $\mathcal{S}_0'= \left\{\calX_{1}^{0}, \dots,  \calX_{n_0'}^{0} \right\}$ to train the classifier, and $\mathcal{S}_0'' = \{\calX^0_{n_0'+1}, \dots, \calX^0_{n_0' + n_0''}\}$ to identify threshold ($n_0' = n''_0$ by default); Class $1$ training data $\mathcal{S}_1= \left\{\calX_{1}^{1}, \dots,  \calX_{n_1}^{1} \right\}$; Define the combined training set $\mathcal{S}:=\mathcal{S}_0'\cup \mathcal {S}_1$
      \item [-] Tucker rank $(r_1,\dots,r_M)$, tolerance parameter $\epsilon$ and maximum number of iterations $T$ for Algorithm~\ref{alg:tensorlda-tucker}
      \item[-] Target level for type I error control $\alpha$, Type I error violation rate $\delta \in (0,1)$
      \end{itemize}}
    \Output{
        \begin{itemize}
            \item Scoring funciton  $\hat{s}^{\text{ LDA}}(\cdot)$ and threshold $\hat{C}^{\text{LDA}}$ for T-LDA classisfier $\bbone\{\hat{s}^{\text{ LDA}}(\cdot) > \hat{C}^{\text{LDA}}\}$
            \item Scoring funciton  $\hat{s}^{\text{ LDA-NP}}(\cdot)$ and threshold $\hat{C}_\alpha^{\text{LDA-NP}}$ for T-LDA-NP classisfier $\bbone\{ \hat{s}^{\text{ LDA-NP}}(\cdot) > \hat{C}_\alpha^{\text{LDA-NP}}\}$.
        \end{itemize}
        }

    \Fn{\FMain{$\mathcal{S}^{\text{input}}$, $(r_1,\dots,r_M)$, $\epsilon$, $T$}}{
    
        Estimate $\hat{\mathcal{B}}^{\text{init}}$ from training set $\mathcal{S}^{\text{input}}$ via \eqref{eqn:lda-discrim-tensor}.

        $\hat{\mathcal{B}} =\text{Algorithm~\ref{alg:tensorlda-tucker}}(\hat{\mathcal{B}}^{\text{init}}, (r_1,...,r_M), \epsilon, T)$\tcp*{Tucker decomposition estimation.}

        $\hat{\mathcal{M}} = \frac{1}{2}(\hat{\mathcal{M}}_0+\hat{\mathcal{M}}_1)$ \tcp*{Estimate the mean tensor for class 0 and class 1.}

        \Return Scoring funciton $\hat{s}(\cdot)=\langle \mathcal{\cdot}, \hat{\mathcal{B}}\rangle$; Threshold $\hat{C} = \langle {\hat{\mathcal{M}}}, \hat{\mathcal{B}}\rangle-\log(n_1/n_0)$.}
    $(\hat{s}^{\text{ LDA}}, \hat{C}^{\text{LDA}}) = \FMain(\mathcal{S} \cup \mathcal{S}_0''
    , (r_1,\dots,r_M), \epsilon, T)$ \tcp*{Merge the reserved class 0 data back to training set, as T-LDA does not need data splitting.}

    $\hat{s}^{\text{ LDA-NP}} = \FMain(\mathcal{S}, (r_1,\dots,r_M), \epsilon, T)$ \tcp*{Get the scoring function from T-LDA.}

    $\hat{C}_\alpha^{\text{LDA-NP}} = \text{Algorithm~\ref{alg:umbrella}}(\mathcal{S}_0'', \hat{s}^{\text{ LDA-NP}}, \alpha, \delta)$ \tcp*{Get the threshold from the NP umbrella algorithm.}

    \Return $(\hat{s}^{\text{ LDA}}, \hat{C}_\alpha^{\text{LDA}})$, $(\hat{s}^{\text{ LDA-NP}}, \hat{C}_\alpha^{\text{LDA-NP}})$
\end{algorithm}

\begin{algorithm}[htpb!]
    \caption{T-NN and T-NN-NP Classifier}
    \label{alg:T-NN-NP}
    \SetKwInOut{Input}{Input}
    \SetKwInOut{Output}{Output}
    \SetKwFunction{FMain}{T-NN}
    \SetKwProg{Fn}{Function}{:}{}

    \Input{ 
      \begin{itemize}
      \item[-]	Randomly split class $0$ training data as $\mathcal{S}_0'= \left\{\calX_{1}^{0}, \dots,  \calX_{n_0'}^{0} \right\}$ to train the classifier, and $\mathcal{S}_0'' = \{\calX^0_{n_0'+1}, \dots, \calX^0_{n_0' + n_0''}\}$ to identify threshold ($n_0' = n''_0$ by default); Class $1$ training data $\mathcal{S}_1= \left\{\calX_{1}^{1}, \dots,  \calX_{n_1}^{1} \right\}$; Define the combined training set $\mathcal{S}:=\mathcal{S}_0'\cup \mathcal {S}_1$
      \item[-] Reserved data $\mathcal{S}^{\text{val}}=\left\{\calX_{1}^{0,\text{val}}, \dots,  \calX_{n_0^{\text{val}}}^{0,\text{val}} \right\} \cup \left\{\calX_{1}^{1,\text{val}}, \dots,  \calX_{n_1^{\text{val}}}^{1,\text{val}} \right\}$ for validation
      \item[ - ] The proposed neural network architecture $F$ (with the max number of epochs $L$).
      \item[-] Target level for type I error control $\alpha$, Type I error violation rate $\delta \in (0,1)$
      \end{itemize}}
    \Output{
        \begin{itemize}
            \item Scoring funciton $\hat{s}^{\text{ NN}}(\cdot)$ for T-NN classisfier $\bbone\{\hat{s}^{\text{ NN}}(\cdot) > 0.5\}$
            \item Scoring funciton  $\hat{s}^{\text{ NN-NP}}(\cdot)$ and threshold $\hat{C}_\alpha^{\text{ NN-NP}}$ for T-NN-NP classisfier $\bbone\{\hat{s}^{\text{ NN-NP}}(\cdot) > \hat{C}_\alpha^{\text{ NN-NP}}\}$.
        \end{itemize}}

    \Fn{\FMain{$\mathcal{S}^{\text{input}}, \mathcal{S}^{\text{val}}, F, L$}}{
    		\For{$i$ in $\{1, \ldots, L\}$\tcp*{For each of the epoch.}}{Train $\hat{s}_{\theta_i}$ by update weights in the neural network $F$ with training dataset $\mathcal{S}^{\text{input}}$, the output is the predicted probability for the observation to be in class 1.

            $\eta_i = \frac{1}{n_{\text{val}}}\sum_{j=1}^{n_{\text{val}}} \bbone\left\{y_j^{\text{val}} = \bbone\{\hat{s}_{\theta_i}(\calX_j^{\text{val}})>0.5\}\right\}$\tcp*{Compute the accuracy of the updated model at current epoch, on validation set.}
    		}
    		$i^{*}\gets $argmax$_{i\in \{1,\ldots,L\}}\eta_i$\tcp*{Stop at the iteration to the best accuracy.}
    	    
            \Return scoring function $\hat{s}(\cdot)=\hat{s}_{\theta_{i^\ast}}(\cdot)$.}

    $ \hat{s}^{\text{ NN}} = \FMain(\mathcal{S} \cup \mathcal{S}_{0}'', \mathcal{S}^{\text{val}}, F, L)$ \tcp*{Merge the reserved data back to training set, as T-NN does not need data splitting.}

    $\hat{s}^{\text{ NN-NP}} = \FMain(\mathcal{S}, \mathcal{S}^{\text{val}}, F, L)$ \tcp*{Get the scoring function from T-NN}
    $\hat{C}_\alpha^{\text{ NN-NP}} = \text{Algorithm~\ref{alg:umbrella}}(\mathcal{S}_0'', \hat{s}^{\text{ NN-NP}}, \alpha, \delta)$ \tcp*{Get the threshold from NP umbrella algorithm.}
    \Return $\hat{s}^{\text{ NN}}$, $(\hat{s}^{\text{ NN-NP}}, \hat{C}_\alpha^{\text{ NN-NP}})$
\end{algorithm}

\section{Theoretical Analysis} \label{sec:theory}

\subsection{Assumptions and Definition}
In this section, we conduct theoretical analysis of T-LDA-NP within the framework of the tensor mixture normal model \eqref{eqn:tgmm2}. We first state the following assumptions.
\begin{assumption}\label{aspt:bounded norm and snr}
    Consider the tensor mixture normal model in \eqref{eqn:tgmm2}.
    \begin{enumerate}[label=(\roman*)]
        \item Let $d_0 = \min_m d_m$, and $r_0 = \min_m r_m$. Suppose there exist a constant $C_0 > 0$, such that $\max_{m}d_m \leq C_0 d_0$, $r_m \leq C_0 r_0$ for any tensor mode $m$, and $C_0^{-1} \leq \lambda_{\min}(\otimes_{m=1}^M \Sigma_m) \leq \lambda_{\max}(\otimes_{m=1}^M \Sigma_m) \leq C_0$. 
        \item $\exists$ positive constants $c$, $C$ such that $c\leq \|\calB\|_F \leq C$ and $\max\{\|\mathcal{M}_0\|_F, \|\mathcal{M}_1\|_F\}\leq C$.
    \end{enumerate}
\end{assumption}
Assumption~\ref{aspt:bounded norm and snr}(i), which requires that the dimensions of each mode be of the same scale, the ranks of each mode not be too large, and the covariance matrices be well-conditioned, is commonly imposed in the tensor literature \citep{zhang2018tensor}.
Assumption~\ref{aspt:bounded norm and snr}(ii) requires the Frobenius norm of $\mathcal{B}$ to be both upper and lower bounded. Asymmetric error control in NP paradigm requires accurate estimation of the decision boundary, which leads to two seemingly contradictory but essential conditions:
\begin{enumerate}[label=(\roman*)]
    \item Sufficiently large signal-to-noise ratio (namely the margin assumption): the majority of samples are far away from the decision boundary, and this is guaranteed by the lower bound on $\|\mathcal{B}\|_F$ in Assumption \ref{aspt:bounded norm and snr}(ii).
    \item Enough information around the decision boundary (namely the detection assumption): the asymmetric error control requires sufficient data near the decision boundary to accurately estimate the threshold, as discussed later in Definition~\ref{def:conditional detection condition}. This is ensured by the upper bound on $\|\mathcal{B}\|_F$ in Assumption \ref{aspt:bounded norm and snr}(ii).
\end{enumerate} 
 While the vector versions of the margin assumption and detection condition have been established as foundations of theoretical analysis in the literature \citep{mammen1999smooth, tong2013plug, tong2020neyman}, we extend these conditions to the tensor setting. Since our analysis concerns asymmetric error control on a high probability event $\mathcal{C}$ (defined in Theorem~\ref{thm:deviation of s}), we define and utilize both conditions to be conditional on $\mathcal{C}$.

\begin{definition}[Conditional margin assumption]\label{def:conditional margin assumption}
A function $f(\cdot)$ satisfies the conditional margin assumption restricted to $\mathcal{C}^\ast$ (a general target set) of order $\bar{\gamma}$ with respect to probability distribution $P_X$ (i.e., $\mathcal{X} \sim P_X$) at level $C^\ast$, if there exists a positive constant $M_0$ such that for any $\iota \geq 0$,
$$
\Pr\bigg\{\big|f(\calX)-C^\ast\big| \leq \iota \mid \calX \in \mathcal{C}^\ast\bigg\} \leq M_0 \iota^{\bar{\gamma}}.
$$
\end{definition}

Definition~\ref{def:conditional margin assumption} extends the margin assumption of \citet{mammen1999smooth} to the conditional setting. Intuitively, this assumption ensures that most data points lie away from the optimal decision boundary.

\begin{definition}[Conditional detection condition]\label{def:conditional detection condition}
A function $f(\cdot)$ satisfies the conditional detection condition restricted to $\mathcal{C}^\ast$ of order $\underline{\gamma}$ with respect to $P_X$ at level $(C^\ast, \delta^\ast)$, if there exists a positive constant $M_1$ such that for any $\iota \in (0, \delta^\ast)$,
\begin{equation*}
\Pr\{C^\ast \leq f(\calX) \leq C^\ast+\iota \mid \calX \in \mathcal{C}^\ast\} \wedge \Pr\{C^\ast-\iota \leq f(\calX) \leq C^\ast \mid \calX \in \mathcal{C}^\ast\} 
\geq M_1 \iota^{\underline{\gamma}}.
\end{equation*}
\end{definition}

In vector setting, the detection condition was introduced by \citet{tong2013plug} and extended to the conditional setting by \citet{tong2020neyman} to counterbalance the margin assumption.

\subsection{Main Results}

In this section, we first establish the estimation error rate for the low-rank discriminant tensor in Proposition~\ref{theorem: tucker}. Building on this error bound, we derive a high-probability bound for the deviation between the estimated and oracle scoring functions in Theorem~\ref{thm:deviation of s}. We then verify that the conditional margin assumption and conditional detection condition hold on this high-probability event, as shown in Propositions~\ref{prop:margin assumption} and~\ref{prop:conditional detection condition}, respectively. These results enable us to prove the NP oracle inequalities for the proposed T-LDA-NP classifier in Theorem~\ref{thm:excess type ii error}.

\begin{proposition}[Tensor LDA]
\label{theorem: tucker} 
Recall $[\Sigma_m]_{m=1}^M$ are the mode-wise covariance matrices of the tensor predictor $\calX$. Denote $\sigma_m = \sigma_{r_m} (\matk(\calB))$. Then 
there exists constants $c$, $C$ and $C_{\rm gap}>0$ which do not depend on $d_m, r_m, \sigma_m$, such that whenever 
\begin{align*}
\sigma_m \geq C_{\rm gap}  \left(\sqrt{\frac{d_m+d_{-m}}{n}} + \frac{\|\matk(\calM_2 - \calM_1)\|_2 \max_{i\le M} d_i}{\sqrt{n d}} \right), m=1,...,M,    
\end{align*}
after sufficient number of iterations in Algorithm \ref{alg:tensorlda-tucker}, the following upper bounds hold with probability at least $1 - n ^{-c} -d^{-c}- \sum_{m=1}^M \exp(-cd_m)$,
\begin{align}
\left\| \hat \calB  -  \calB \right\|_{\rm F} & \le C \sqrt{ \frac{r+\sum_{m=1}^M d_m r_m}{n  }  }  + C\|\cM_2-\cM_1 \|_{\rm F} \cdot \frac{\max_{1\le m\le M}d_m}{\sqrt{n  d}} . \label{eqn:lda-b-tucker} 
\end{align}
\end{proposition}



The estimated classifier is obtained by plugging in the low-rank estimate $\hat{\calB}$: $\hat{\phi}^{\text{ LDA-NP}}_\alpha(\calX) =  \bbone\{\hat{s}^{\text{ LDA-NP}}(\calX) > \hat{C}_\alpha^{\text{LDA-NP}}\}$, where $\hat{s}^{\text{ LDA-NP}}(\calX) = \langle \calX, \hat{\calB} \rangle$ and $\hat{C}_\alpha^{\text{LDA-NP}}$ is the threshold selected by the NP umbrella algorithm (Algorithm~\ref{alg:umbrella}). Our goal is to establish the NP oracle inequalities, which consist of two parts: first, the type I error $R_0(\hat{\phi}^{\text{ LDA-NP}}_\alpha)$ is bounded above by $\alpha$ (already shown in Proposition~\ref{prop:umbrella}), and second, the excess type II error $R_1(\hat{\phi}^{\text{ LDA-NP}}_\alpha) - R_1(\phi_\alpha^\ast)$ vanishes as the sample size increases. We establish both the type I error control and the vanishing excess type II error on a high-probability set, which is stronger than control in expectation. For convenience, we denote $n_{\min} = n_0 \wedge n_1$ in the following analysis.

Since controlling the excess type II error requires accurate estimation of the scoring function, we first establish a high-probability bound for the deviation between $\hat{s}^{\text{ LDA-NP}}$ and $s^\ast$.

\begin{theorem}[High-probability bound for scoring function deviation]
    \label{thm:deviation of s}
    Suppose Assumption~\ref{aspt:bounded norm and snr} and the assumptions of Proposition~\ref{theorem: tucker} hold. Define the event
    \[
    \mathcal{C}=\mathcal{C}(t) = \left\{\calX: \left| \langle\calX-\calM, \hat{\calB}-\calB\rangle\right|/(\sqrt{\|\bSigma_v\|}\|\hat{\calB}-\calB\|_F) < t \right\},
    \]
    where $\bSigma_v = \Sigma_M \otimes \ldots \otimes \Sigma_1$. Then,
    \begin{enumerate}[label=(\roman*)]
    \item $\mathcal{C}(t)$ has high probability under both $\mathbb{P}_0$ and $\mathbb{P}_1$, where $\mathbb{P}_i$ denotes the distribution of $\calX$ given $Y=i$ for $i=0,1$:
    \[
    \mathbb{P}_i(\calX \notin \mathcal{C}(t)) \leq  2 \exp(-t^2/2).
    \]
    \item On $\mathcal{C}(t)$, the deviation between $\hat{s}^{\text{ LDA-NP}}(\mathcal{X})$ and $s^\ast(\mathcal{X})$ is bounded by
    \[
    \| s^\ast - \hat{s}^{\text{ LDA-NP}} \|_{\infty, \mathcal{C}} := \max_{\mathcal{X}\in\mathcal{C}} |s^\ast(\calX) - \hat{s}^{\text{ LDA-NP}}(\calX)| \lesssim t\sqrt{\frac{d_0 r_0}{n_{\min} } },
    \]
    where $d_0 = \min_m d_m$ and $r_0 = \min_m r_m$.
    \end{enumerate}
 \end{theorem}
The goal of Theorem~\ref{thm:deviation of s} is to provide a high-probability bound on the deviation between the estimated and the oracle scoring functions. Intuitively, achieving control with higher probability requires tolerating a larger deviation, calibrated by the parameter \(t\).

 \begin{remark}\label{rmk:deviation of s}    
    The asymptotic behavior of the deviation bound depends on how $t$ and $d_0$ scale with the sample size. Following the convention in Lemma~10 of \citet{tong2020neyman}, we set $t$ to be in polynomial order of the sample size, i.e. $t \asymp n_{\min}^a$ for some positive constant $a$. To obtain meaningful scaling, we assume $d_0 \asymp n_{\min}^b$ for some positive constant $b$ and that $r_0$ remains of constant order. This yields
        \[
        \PP_i(\calX \notin \mathcal{C}) \leq 2 \exp(-t^2/2)\asymp \exp(-n_{\min}^{2a})\quad\text{for }i=0,1,
        \]
        and
        \[ 
        \| s^\ast - \hat{s}^{\text{ LDA-NP}} \|_{\infty, \mathcal{C}}  \lesssim n_{\min}^{a+\frac{1}{2}b-\frac{1}{2}}. 
        \]
        For the subsequent theoretical analysis, it suffices to have $a + \frac{b}{2} < \frac{1}{2}$ with $a, b > 0$. For instance, choosing $a = 1/3$ and $b = 1/6$ gives
        \[
        \PP_i(\calX \notin \mathcal{C})  \asymp \exp(-n_{\min})\quad\text{for }i=0,1, \qquad \text{and} \qquad  \| s^\ast - \hat{s}^{\text{ LDA-NP}} \|_{\infty, \mathcal{C}}  \asymp n_{\min}^{-\frac{1}{12}}.
        \]
    
 \end{remark}

Since our analysis takes place on a high-probability set, we must verify that the conditional margin assumption and conditional detection condition hold on the set $\mathcal{C}$ defined in Theorem~\ref{thm:deviation of s}. For the conditional margin assumption (Definition~\ref{def:conditional margin assumption}), Proposition~\ref{prop:margin assumption} establishes, via Bayes' theorem, that the conditional margin assumption is weaker than its unconditional counterpart.

\begin{proposition}[Conditional margin assumption]
    \label{prop:margin assumption}
    Under Assumption~\ref{aspt:bounded norm and snr} and the assumptions of Proposition~\ref{theorem: tucker}, the conditional margin assumption holds with order $\bar{\gamma}=1$ for the function $s^\ast$ on the set $\mathcal{C}$ with respect to $\PP_0$ at level $C_\alpha^{\ast\ast}$.
\end{proposition}

Unlike the conditional margin assumption, the conditional detection condition (Definition~\ref{def:conditional detection condition}) is stronger than its unconditional counterpart. While \citet{tong2020neyman} verified this condition for specific examples, prior research has not identified general sufficient conditions for its validity. As one contribution of this paper, we establish such general results in Proposition~\ref{prop:conditional detection condition} by first applying Bayes' theorem to reverse the conditional probability, then constructing a subset of $\mathcal{C}$ that is independent of the conditioning event to lower bound the conditional probability.

\begin{proposition}[Conditional detection condition]
    \label{prop:conditional detection condition}
    Under Assumption~\ref{aspt:bounded norm and snr} and the assumptions of Proposition~\ref{theorem: tucker}, let $C_B$ and $C_L$ denote the upper bound of $\|\calB\|_F$ and the lower bound of the signal-to-noise ratio, respectively. In other words, $\|\calB\|_F \leq C_B$ and $\vect(\cD)^\top (\bSigma_v^{-1}) \vect(\cD) \geq C_L$. If
    {\footnotesize
    \begin{equation*}
    t\geq \max\left\{\Phi^{-1}(1-\alpha), C_0^M\sqrt{1+3\frac{C_B}{C_L}}\Phi^{-1}\left(\frac{3}{4}\right)+\frac{C_0^{3M/2}C_B}{C_L}\left(\delta^*+\sqrt{C_L}\Phi^{-1}(1-\alpha)+C_L/2\right)\right\},
    \end{equation*}}
    where $C_0$ is from Assumption~\ref{aspt:bounded norm and snr} and $\Phi^{-1}(\cdot)$ is the quantile function of the standard normal distribution, then the conditional detection condition holds with order $\underline{\gamma}=1$ for the function $s^\ast$ on the set $\mathcal{C}$ with respect to $\PP_0$ at level $(C_\alpha^{\ast\ast}, \delta^*)$ for any $\delta^*$ satisfying
    \[\delta^* \in (0, t(\sqrt{\norm{\bSigma_v}}\norm{\calB}_F)-|\langle \cM,\calB \rangle-C_\alpha^{**}|).\]

\end{proposition}

We now present the main theoretical result, Theorem~\ref{thm:excess type ii error}, which establishes the NP oracle inequalities for the proposed T-LDA-NP classifier.

\begin{theorem}[NP oracle inequalities]
    \label{thm:excess type ii error}
    Suppose Assumption~\ref{aspt:bounded norm and snr} and the assumptions of Proposition~\ref{theorem: tucker} are satisfied. For any arbitrary $\delta_0' \in (0,1)$, suppose the size of the holdout class 0 data satisfies
    $n_0'' \geq \max \{\delta_0^{-2}, \delta_0^{\prime-2},4 /(\alpha \delta_0), 10^{4}M_1(\delta^{\ast})^{-4\underline{\gamma}}\}$,  where $\delta_0$ is the specified violation rate, and the constants $C_0$, $C_B$, $C_L$, $\delta^\ast$, $M_1$, and $\underline{\gamma}$ are as defined in Assumption~\ref{aspt:bounded norm and snr}, Definition~\ref{def:conditional detection condition} and Proposition~\ref{prop:conditional detection condition}. For any sufficiently large $t$ from Theorem~\ref{thm:deviation of s} satisfying
    \[\begin{aligned}
    t \geq \max\Bigg\{& C_0^M\sqrt{1+3C_B/C_L}\Phi^{-1}(3/4)+\frac{C_0^{3M/2}C_B}{C_L}\left(\delta^\ast+\sqrt{C_L}\Phi^{-1}(1-\alpha)+C_L/2\right),\\
    &\Phi^{-1}(1-\alpha),\quad \sqrt{-2\log\left( \frac{1}{8}M_1(\delta^{*})^{\underline{\gamma}} \right)}\Bigg\},
    \end{aligned}\]
    the following hold with probability at least $1-\delta_0-\delta_0^{\prime} - C\exp(-cd_0)$:
    \begin{enumerate}[label=(\roman*)]
    \item The type I error satisfies 
    \begin{equation}
    \PP_0\left(\hat{s}^{\text{ LDA-NP}}(\calX)\geq\hat{C}_\alpha^{\text{LDA-NP}} \right)<\alpha.
    \end{equation}
    \item The excess type II error satisfies
    \begin{equation} \label{thmnp:type2}
    \begin{aligned}
    & \PP_1\left( \hat{s}^{\text{ LDA-NP}}(\calX)<\hat{C}_\alpha^{\text{LDA-NP}} \right)-\PP_1\left( s^\ast(\calX)<C_\alpha^{\ast\ast} \right) \\
    \leq& C_1'' ({n_0''})^{-\min\left\{\frac{1+\bar{\gamma}}{4\underline{\gamma}}, \frac{1+\bar{\gamma}}{4\bar{\gamma}}, \frac{1}{4} \right\}}
    + C_2''e^{-\frac{t^2}{2}\min\left\{\frac{1+\bar{\gamma}}{\underline{\gamma}}, \frac{1+\bar{\gamma}}{\bar{\gamma}}, 1\right\}}
    +C_3'' \left( t\sqrt{\frac{d_0 r_0}{n_{\min}}} \right)^{1+\bar{\gamma}}.
    \end{aligned}
    \end{equation}
    \end{enumerate}
\end{theorem}
\begin{remark}
    Part (i) ensures that the type I error is controlled at the pre-specified level $\alpha$ with high probability. Part (ii) provides an upper bound for the excess type II error, which decomposes into three terms. The first term arises from threshold estimation in the umbrella algorithm and decreases with the holdout class 0 sample size $n_0''$. The second and third terms capture the scoring function estimation error: the second term represents the probability mass outside the high-probability set, while the third term quantifies the deviation between the estimated and oracle scoring functions within this set. As $t$ increases, the second term diminishes exponentially, while the third term decreases with sample size growth, provided $t$ and $d_0$ scale appropriately.

    As discussed in Remark~\ref{rmk:deviation of s}, when $t \asymp n_{\min}^{\frac{1}{3}}$, $d_0 \asymp n_{\min}^{\frac{1}{6}}$ and $\bar{\gamma} = \underline{\gamma} =1$ (by Propositions \ref{prop:margin assumption} and \ref{prop:conditional detection condition}), the upper bound in \eqref{thmnp:type2} simplifies to
    \[
    C_1'' ({n_0''})^{- \frac{1}{4} }
    + C_2''\exp \left\{-\frac{1}{2}n_{\min}^{\frac{2}{3}}\right\}
    +C_3'' n_{\min}^{-\frac{1}{6}},
    \]
    which vanishes as the sample size grows. More generally, for any $0 < b < 1$ such that $d_0 \asymp n_{\min}^b$, choosing $0 < a < \frac{1-b}{2}$ ensures diminishing excess type II errors.
    
\end{remark}
    
\section{Simulation}\label{sec:simulation}

In this section, we evaluate the performance of the proposed NP classifiers, T-NN-NP and T-LDA-NP through simulation studies and compare with their counterparts T-NN and T-LDA, designed for minimizing the overall classification error, together with the vectorized V-LDA. We assess performance in terms of type I error, type II error, accuracy, and violation rate.

Example \ref{exp:vary train size} provides the main illustration of the asymptotic behavior of the proposed procedure. By varying the training sample size, it verifies the theoretical type I error control and highlights the gains from leveraging tensor structure rather than relying on vectorized representations. Examples \ref{exp:vary shape} and \ref{exp:vary rank} further assess the benefits of the low-rank tensor model: Example \ref{exp:vary shape} investigates robustness as tensor dimensions grow, while Example \ref{exp:vary rank} evaluates robustness to misspecification of the Tucker rank supplied to the T-LDA classifiers. The type I error control remains valid and accuracy is largely unaffected in both cases. Additionally, Example \ref{exp:vary distribution} in Supplementary Material evaluates the robustness of type I error control under non-TGMM (i.e., not tensor normal) distributions.

Across all examples, we draw $n_0^{\text{train}}$ class~0 and $n_1^{\text{train}}$ class~1 observations for training, reserving half of the class~0 samples for threshold estimation in T-NN-NP and T-LDA-NP. For testing, we independently draw \(n_0^{\text{test}}\) class~0 and \(n_1^{\text{test}}\) class~1 observations. Unless otherwise noted, the experiments are repeated for 500 times and use the following settings with $\alpha = 0.05$, $\delta =0.1$:

\begin{itemize}
    \item Let $\cM_0=\mathbf{0}$ and $\bSigma_m = I_{d_m}$ for all $m \in [M]$. The discriminant tensor $\calB$ is randomly generated with Tucker rank $(4,6,3)$ satisfying $\text{SNR} = \sqrt{\langle \calB,\calB \rangle} = \|\calB\|_F = 7$. Then $\calB = \cD \times_{m=1}^M \Sigma_m^{-1} = \cD = \cM_1$.
    \item The class size ratio $\eta := n_1^{\text{train}}/n_0^{\text{train}} = n_1^{\text{test}}/n_0^{\text{test}}$.  In particular, $\eta=1$ represents the balanced-classes setting. The total test sample size is fixed at $60000$.
\end{itemize} 



\begin{figure}[htbp]
    \centering
    \includegraphics[width=0.95\linewidth]{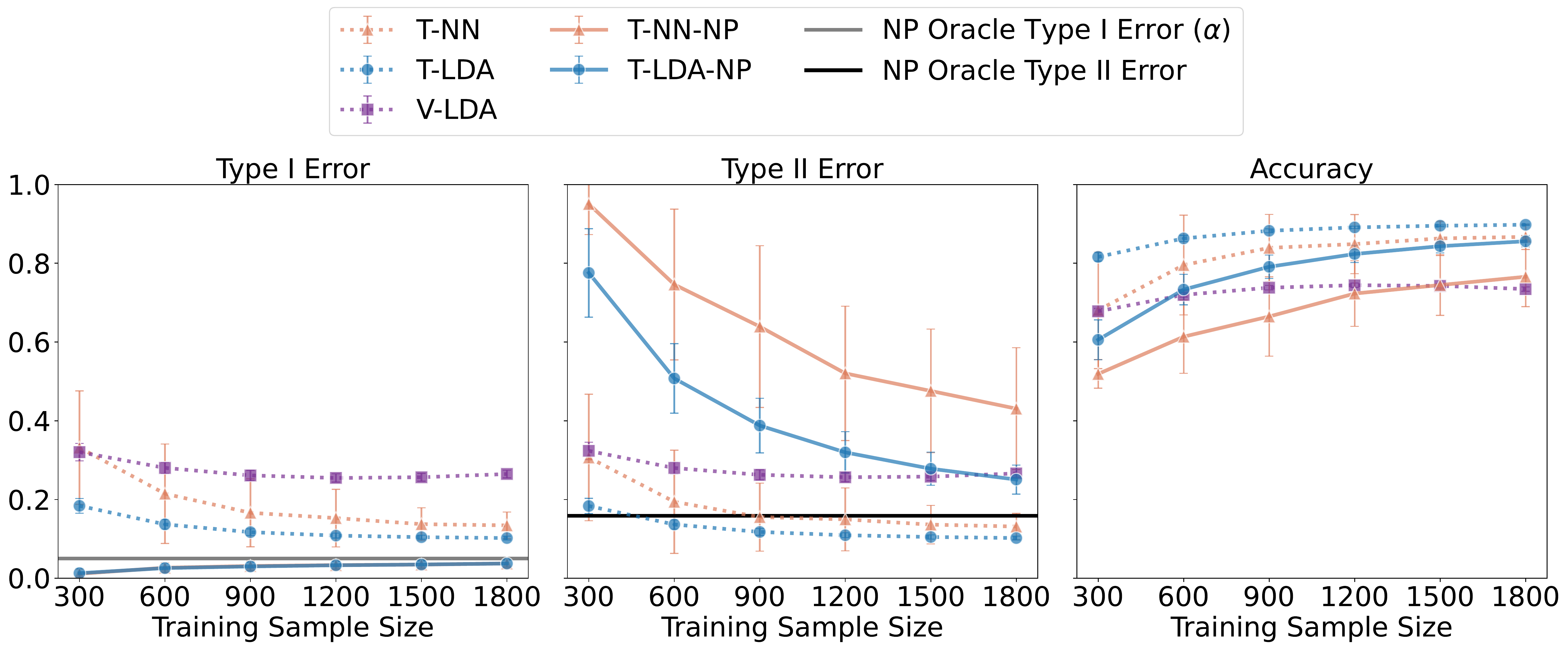}
    \caption{Average Type I Error, Average Type II Error and Accuracy in Example \ref{exp:vary train size} ($\alpha=0.05, \delta=0.1$), comparing methods with increasing size of training data.
    }
    \label{fig:vary train size}
\end{figure}

\begin{example}\label{exp:vary train size}
    We vary the total training sample size in $\{300, 600, 900, 1200, 1500, 1800\}$ with $\eta=1$ to validate the theoretic results of the controlled type I error and the diminishing excess type II error in T-NN-NP and T-LDA-NP. All the data are generated from tensor normal distribution (e.g. TGMM) with shape $(15,15,15)$. The average type I error, type II error and accuracy are reported in FIG \ref{fig:vary train size} and the violation rate is summarized in Table \ref{tbl:vary train size}.
\end{example}

\begin{table}[h!]
    \caption{Violation rates in Example \ref{exp:vary train size}: $\alpha=0.05, \delta=0.1$, comparing methods with increasing size of training data.}
    \label{tbl:vary train size}
    \centering
    \begin{tabular}{|c||c|c|c|c|c|}
    \hline
    $n^{\text{train}}$ & T-NN & T-LDA & V-LDA & T-NN-NP & T-LDA-NP\\
    \hline
    300 & 1.000 & 1.000 & 1.000 & .014 & .020\\
    \hline
    600 & 1.000 & 1.000 & 1.000 & .054 & .030\\
    \hline
    900 & 1.000 & 1.000 & 1.000 & .086 & .040\\
    \hline
    1200 & 1.000 & 1.000 & 1.000 & .066 & .060\\
    \hline
    1500 & 1.000 & 1.000 & 1.000 & .054 & .048\\
    \hline
    1800 & 1.000 & 1.000 & 1.000 & .096 & .082\\
    \hline
    \end{tabular}

    \end{table}

The impact of asymmetric error control is clear in FIG \ref{fig:vary train size} and Table \ref{tbl:vary train size}. For the NP procedures T-NN-NP and T-LDA-NP, the average type I errors are effectively controlled under $\alpha = 0.05$, with violation rates consistently below $\delta = 0.1$. In contrast, for their counterparts T-NN, T-LDA and V-LDA, all violation rates are equal to 1 indicating that the type I errors always exceed $\alpha$. Furthermore, as the sample size increases, the type I errors of T-LDA-NP and T-NN-NP approach $\alpha$ from below, reflecting increasingly precise classifiers with diminishing excess type II errors and validating Theorem \ref{thm:excess type ii error}. While T-NN and T-LDA also show improved accuracy with larger sample sizes, as noted earlier, they fail to control type I errors. V-LDA performs the worst, as it disregards the underlying tensor structure. Comparing LDA-based with NN-based classifiers, as the underlying model is TGMM, it is not surprising to see that the former achieve smaller and more rapidly diminishing excess type II errors and higher accuracy.

Intuitively, increasing the ratio $\eta$ leads to a relatively smaller class $0$ and thus tends to exacerbate the type I error. We therefore also consider an imbalanced setting with $\eta = 2$, with similar results and interpretations reported in FIG~\ref{fig:vary train size imbalanced} and Table~\ref{tbl:vary train size imbalanced} in the Supplementary Material. 

\begin{example}
    \label{exp:vary shape}
    In this experiment, we assess the effect of tensor shape by fixing the aspect ratio \(d_1 : d_2 : d_3 = 1 : 1 : 1\) and varying the common dimension \(d=d_1\) (in the other examples, we set $d = 15$). We consider \(d \in \{13,14,15,16,17,18\}\) and generate data from TGMM.  The resulting average type~I/II errors and accuracy are reported in FIG~\ref{fig:vary shape}, and the violation rates are summarized in Table~\ref{tbl:vary shape}, both in Supplementary Material.

\end{example}

It is worth noting that although $d$ varies only over a small range, the total number of entries in the 3-way tensor increases substantially: from \(13^3\) to \(18^3\). The main patterns in FIG~\ref{fig:vary shape} and Table~\ref{tbl:vary shape} are consistent with those in Example~\ref{exp:vary train size}. Both T-NN-NP and T-LDA-NP efficiently maintain type I errors below $\alpha = 0.05$ and violation rates below $\delta = 0.1$. Although one might expect performance to deteriorate as $d$ increases, all methods exhibit little sensitivity to the growing dimension, as indicated by the nearly unchanged average type I error, average type II error, and accuracy. For T-LDA and T-LDA-NP in particular, this robustness highlights the effectiveness of the low-rank Tucker structure: the Tucker rank is fixed at $(4,6,3)$ while the ambient tensor dimensions increase, and the convergence rate in Proposition~\ref{theorem: tucker} depends on both the rank and the mode dimensions, so the resulting change in performance is minor.

\begin{example}\label{exp:vary rank}
    In this experiment, we evaluate the robustness of the proposed methods to misspecification of the Tucker rank used by T-LDA and T-LDA-NP. In practice, the true Tucker rank is unknown and must be estimated, making such misspecification unavoidable. All data are generated from a TGMM with tensor shape \((15,15,15)\), where the true low-rank tensor \(\mathcal{B}\) has Tucker rank $(4,\,6,\,3)$. During estimation, however, we intentionally use misspecified ranks of the form \((4,\, 6,\, 3) + \Delta \), with $\Delta \in \{(0,\,0,\,0),\, (2,\,0,\,0),\, (-2,\,0,\,0),\, \\ (0,\,2,\,0),\, (0,\,-2,\,0),\, (0,\,0,\,2),\, (0,\,0,\,-2)\}$.
 The resulting average accuracies, type~I errors, type~II errors and violation rates are summarized in {Tables~\ref{tbl:varyrank_accuracy}-\ref{tbl:varyrank_violation_rate}}, respectively.
\end{example}
Across all considered working Tucker ranks, the type~II errors and violation rates for both T-LDA and T-LDA-NP vary only slightly, demonstrating robustness to moderate rank misspecification. Since the chosen Tucker rank does not affect T-NN or T-NN-NP, we report results only for the LDA-based methods. Notably, nearly all violation rates for T-LDA-NP remain below $\delta=0.1$.

\section{Real Data Analysis}\label{sec:application}

In modern computational chemistry and molecular biology, accurately characterizing the structure and function of small molecules and proteins is a problem of broad scientific and societal importance. Molecular structure analysis underpins key applications such as early-stage drug discovery, enzyme function prediction, mutagenicity or carcinogenicity screening, and the design of new chemical compounds and materials. Decisions at this stage determine which compounds enter costly downstream pipelines, including \textit{in vitro} assays, \textit{in vivo} toxicity tests, and clinical investigations, with direct consequences for both public health and biomedical innovation.

Despite this importance, learning from molecular and protein structures remains highly challenging. These objects possess complex, irregular, and multi-scale geometry, with chemical and biological properties encoded in both local connectivity (e.g., bonds or residue adjacency) and global structural features (motif organization, spatial folding). To capture these high-order interactions, we transform molecular graphs into topological descriptors and encode them into higher-order tensors, which can be further modeled by flexible tensor neural networks (T-NN) capable of representing nonlinear decision boundaries.

In classification tasks, a further challenge arises from asymmetric classification risk. In many biochemical screening tasks, type~I and type~II errors carry substantially different costs, and the class sizes are markedly imbalanced. For example, misclassifying a mutagenic compound as safe (type~I error) poses significant health risks and regulatory consequences, whereas misclassifying a non-mutagenic compound as mutagenic (type~II error) triggers unnecessary and expensive follow-up experiments. Similar considerations apply to carcinogenicity prediction (PTC\_MM dataset) and biological activities classification (COX2 and BZR datasets): missing active compounds (type~I error) may obscure crucial biochemical insights and potential pharmaceutical ingredient in drug discovery process, while false signals waste resources. These asymmetries call for explicit, finite-sample control of the more costly type~I error, a guarantee provided by our T-NN-NP procedure.

We analyze four biochemical datasets from \cite{Morris+2020}, MUTAG, COX2, BZR and PTC\_MM, to showcase how tensor-based modeling  with NP-type error control addresses challenges arising from complex molecular structures, imbalanced classes, and asymmetric classification risks. For brevity, the full analyses of PTC\_MM are deferred to Supplementary Material. These datasets consist of small-molecule graphs, where nodes represent atoms and edges encode chemical bonds, enriched with information on atom types, bond multiplicities, and other chemical properties \citep{Morris+2020}. Following \cite{chazal2021introduction, adams2017persistence}, we first convert each graph into persistence diagrams via sublevel-set filtrations and then into persistence images, whose stacking across filtrations yields tensor representations that retain multiscale topological information for tensor classification.

For all experiments, we randomly split the data into training and testing sets 100 times and summarize the results. Because these datasets violate the Gaussian assumptions required by T-LDA and T-LDA-NP, our analysis focuses on T-NN and T-NN-NP. In the following subsections, we describe the analysis and results for each dataset.

\subsection{Chemical Mutagenicity (MUTAG)}
The MUTAG dataset contains 188 chemical compounds classified by their mutagenic effect on \textit{S.\ typhimurium} TA98. The data were originally collected by \cite{debnath1991structure}, with mutagenicity quantified using the log-transformed TA98 response (revertants/nmol) from the Ames test. In this study, we adopt the binary labels from \cite{Morris+2020}, labeling compounds with \emph{high mutagenicity} as class~0 to reflect the practical priority of reducing potential health risks. This yields $n_0 = 125$ high-mutagenicity compounds (class~0), and $n_1 = 63$ low-mutagenicity compounds (class~1).

{
\begin{table}[ht]\tiny
\caption{Results on the MUTAG dataset.}
\centering
\begin{tabular}{|c||c|c|c|c|c|c|c|c|}
\hline
 & \multicolumn{2}{c|}{$\alpha = 0.03$, $\delta = 0.40$} & \multicolumn{2}{c|}{$\alpha = 0.03$, $\delta = 0.50$} & \multicolumn{2}{c|}{$\alpha = 0.05$, $\delta = 0.50$} & \multicolumn{2}{c|}{$\alpha = 0.07$, $\delta = 0.50$} \\
\hline
Algorithm & T-NN & T-NN-NP & T-NN & T-NN-NP & T-NN & T-NN-NP & T-NN & T-NN-NP \\
\hline
Type II Error & .190(.151) & .845(.204) & .190(.151) & .786(.239) & .190(.151) & .713(.224) & .190(.151) & .613(.237) \\
\hline
Accuracy & .870(.055) & .676(.075) & .870(.055) & .713(.076) & .870(.055) & .718(.082) & .870(.055) & .746(.081) \\
\hline
Violation Rate & .859 & .404 & .859 & .340 & .707 & .270 & .707 & .360 \\
\hline
\end{tabular}\label{tb:MUTAG}
\end{table}}

Table~\ref{tb:MUTAG} reports the average type~II error, overall accuracy, and the type~I error violation rates across four $(\alpha,\delta)$ configurations. Several patterns emerge. First, T-NN exhibits a type I error of 0.099 and a violation rate exceeding 70\% in all configurations, indicating that its type~I error is substantially inflated and cannot be controlled at the targeted levels $\alpha=0.03$ or $\alpha=0.05$. This strongly motivates the NP-adjusted T-NN-NP procedure. Second, fixing $\alpha = 0.03$ and increasing the allowed violation level from $\delta=0.4$ to $\delta=0.5$ relaxes the type~I error constraint. As expected, this leads to a reduction in the average type~II error from $0.845$ to $0.786$ and improves overall accuracy from $0.676$ to $0.713$. Third, holding $\delta=0.5$ constant while increasing $\alpha$ from $0.03$ to $0.07$ further eases the restriction on type~I error. This yields a monotonic decrease in type~II error to $0.613$, and an accuracy increase to $0.746$. Although the accuracy of T-NN-NP remains slightly below that of T-NN, this reflects the inevitable trade-off when enforcing stringent control of type~I error, which is prioritized in this application. Finally, across all configurations, the empirical violation rates of T-NN-NP remain at or below the targeted level~$\delta$ (the sole exception being $0.404$, effectively equal to $0.4$)\footnote{It is important to note that the reported type~I error violation rate is an empirical surrogate rather than the exact theoretical quantity. The estimate deviates from the ideal definition in two ways: 1) within each repetition, the population type~I error is inferred from its empirical counterpart computed on a small test set; 2) the violation rate itself is a long-run probability defined over infinitely many repetitions, yet our estimate derives from finite experimental runs. 
These discrepancies are unavoidable in computational studies and represent the standard practical approach to evaluating error control.
}, confirming that the NP procedure successfully delivers the intended error control.

\subsection{Cyclooxygenase-2 Inhibitor (COX2)}
The COX2 dataset contains 467 cyclooxygenase-2 inhibitors classified by their \textit{in vitro} inhibition. The data were originally published by \cite{sutherland2003spline}, with activity quantified using IC50 measuring inhibition of human recombinant enzyme. We adopt the binary labels from \cite{Morris+2020}, labeling compounds with \emph{high activity} as class~0 to prioritize identifying potent candidates for discovery of drug including anti-inflammatory agents and pain relievers. This yields $n_0 = 365$ high-activity compounds (class~0), and $n_1 = 102$ low-activity compounds (class~1).

\begin{table}[h!]
\centering
\caption{Results on the COX2 dataset.}
\begin{tabular}{|c||c|c|c|c|c|c|}
\hline
 & \multicolumn{2}{c|}{$\alpha = 0.01$, $\delta = 0.30$} & \multicolumn{2}{c|}{$\alpha = 0.01$, $\delta = 0.40$} & \multicolumn{2}{c|}{$\alpha = 0.01$, $\delta = 0.50$} \\
\hline
Algorithm & T-NN & T-NN-NP & T-NN & T-NN-NP & T-NN & T-NN-NP \\
\hline
Type II Error & .783(.117) & .955(.066) & .783(.117) & .939(.066) & .783(.117) & .943(.059) \\
\hline
Accuracy & .808(.019) & .783(.013) & .808(.019) & .786(.014) & .808(.019) & .786(.013) \\
\hline
Violation Rate & .67 & .26 & .67 & .29 & .67 & .25 \\
\hline
\end{tabular}\label{tb:COX2}
\end{table}

Table~\ref{tb:COX2} reports type~II error, overall accuracy, and type~I error violation rates across three $(\alpha,\delta)$ configurations with fixed $\alpha = 0.01$ to reduce missed active compounds. The standard T-NN yields a violation rate of $0.67$ across all settings, demonstrating that type~I error control fails substantially at the nominal level $\alpha$. With T-NN-NP, empirical violation rates drop to between $0.25$ and $0.29$, consistently below their respective target thresholds $\delta$ and confirming effective type~I error control in practice. The corresponding accuracy of T-NN-NP ranges from $0.783$ to $0.786$, representing a modest loss relative to T-NN's $0.808$ accuracy, and reflecting an unavoidable trade-off for achieving the stringent type I error control required in drug discovery. Varying $\delta$ from $0.3$ to $0.5$ produces negligible performance changes here. This insensitivity arises from the interplay between the limited left-out class $0$ sample size (e.g., 182), and the stringent $\alpha = 0.01$: under these conditions, the order statistic-based nonparametric threshold remains insensitive to $\delta$ in the examined range.

\subsection{Benzodiazepine Receptor Binding (BZR)}

The BZR dataset \citep{sutherland2003spline} contains 405 ligands for the benzodiazepine receptor classified by their \textit{in vitro} binding affinities, quantified using IC50 values measuring inhibition of $[^3H]$ diazepam binding. Similar to COX2, we label compounds with \emph{high activity} as class~0 to prioritize identifying potent candidates for drug discovery including anti-anxiety agents \cite{Morris+2020}. This yields $n_0 = 319$ high-activity compounds (class~0), and $n_1 = 86$ low-activity compounds (class~1).

\begin{table}[h!]
\centering
\caption{Results on the BZR dataset.}
\begin{tabular}{|c||c|c|c|c|c|c|}
\hline
 & \multicolumn{2}{c|}{$\alpha = 0.01$, $\delta = 0.30$} & \multicolumn{2}{c|}{$\alpha = 0.01$, $\delta = 0.40$} & \multicolumn{2}{c|}{$\alpha = 0.01$, $\delta = 0.50$} \\
\hline
Algorithm & T-NN & T-NN-NP & T-NN & T-NN-NP & T-NN & T-NN-NP \\
\hline
Type II Error & .552(.151) & .836(.111) & .552(.151) & .864(.121) & .552(.151) & .864(.127) \\
\hline
Accuracy & .858(.027) & .819(.021) & .858(.027) & .814(.024) & .858(.027) & .813(.024) \\
\hline
Violation Rate & .8 & .33 & .8 & .25 & .8 & .34 \\
\hline
\end{tabular}\label{tb:BZR}
\end{table}

The results in Table~\ref{tb:BZR} exhibit patterns broadly consistent with COX2. T-NN-NP successfully controls the type~I error with empirical violation rates ranging from $0.25$ to $0.34$, all below their corresponding target levels $\delta$ except for one instance of $0.34$ (effectively equal to $\delta=0.3$; cf. MUTAG footnote reasoning) and hence the high-probability type I error control is practically achieved. As a trade-off, T-NN-NP accuracy ranges from $0.813$ to $0.819$, a modest sacrifice compared to T-NN's $0.858$. 

\section{Discussion and Conclusion} \label{sec:conclusion}

In this work, we implement the NP framework within tensor-based classification models, including neural networks and binary TGMM, proposing two new classifiers, T-NN-NP and T-LDA-NP. Both methods provide finite-sample, high-probability control of type~I error, a property that is essential in scientific applications where the consequences of type~I error outweigh those of type~II error. We further show that the conditional detection condition holds under mild assumptions, a fact that has not been shown previously, and establish that the excess type~II error for T-LDA-NP vanishes asymptotically.

The empirical studies on biochemical datasets highlight the practical relevance of our contributions. While we focused on molecular mutagenicity, carcinogenicity, and enzyme activity prediction in this paper, our methodology has the potential to extend to numerous other domains, including medical imaging, spatiotemporal climate data, neuroimaging connectomes, multiway sensor arrays, and recommendation systems. In many of these fields, domain-specific constraints also play a critical role. As such, an exciting direction for future work is to incorporate alternative forms of constraints into tensor classification. Examples include fairness constraints (demographic parity, equalized odds), robustness constraints against adversarial perturbations, and cost-sensitive constraints for industrial decision systems could all be integrated into an NP-style framework. Another methodological extension involves enriching the structure of tensor Gaussian mixture models: allowing heterogeneity in $\boldsymbol{\Sigma}_k$ (i.e., tensor quadratic discriminant analysis) or modeling task-specific covariance structures may improve expressiveness and interpretability. Coupling such extensions with NP error control would broaden the applicability of tensor-based classifiers in settings where structured, asymmetric risks are fundamental considerations.


%
%
\bibliographystyle{\mybibsty}
\bibliography{main_arXiv}

\clearpage
\setcounter{page}{1}
\begin{center}
{\Large Supplementary Material of ``\TITLE''}\\
\author{
    Lingchong~Liu$^\sharp$ \hspace{8ex}
    Elynn~Chen$^\dag$ \hspace{8ex}
    Yuefeng~Han$^\flat$ \hspace{8ex}
    Lucy~Xia$^\sharp$  \\ \normalsize
    $^{\sharp}$The Hong Kong University of Science and Technology  \\
    $^\dag$New York University \hspace{8ex}
    $^\flat$ University of Notre Dame \hspace{8ex}
    }
\end{center}

\renewcommand{\thealgocf}{S\arabic{algocf}}
\renewcommand{\thefigure}{S\arabic{figure}}
\renewcommand{\thetable}{S\arabic{table}}
\renewcommand{\theequation}{S\arabic{equation}}
\renewcommand{\theexample}{S\arabic{example}}
\renewcommand{\thesection}{S\arabic{section}}

\setcounter{figure}{0}
\setcounter{table}{0}
\setcounter{equation}{0}
\setcounter{example}{0}
\setcounter{section}{0}
\setcounter{algocf}{0}

This supplementary material includes the NP umbrella algorithm, additional numerical results in simulation and real data applications, and proofs of main results and technical lemmas.

\section{NP Umbrella Algorithm}

The umbrella algorithm proposed by \cite{tong2018neyman} is a general sample-splitting-based algorithm to construct a classifier that satisfies the NP type~I error control, summarized in Algorithm~\ref{alg:umbrella}. 

\begin{algorithm}[h!]
    \SetKwInOut{Input}{Input}
    \SetKwInOut{Output}{Output}
    \Input{
        \begin{itemize}
            \item[-] Hold-out class 0 training data $S_0''$ with size $n_0'''$
            \item[-] Trained scoring function $\hat{s}$ 
            \item[-] Type I error upper bound $\alpha$ ($\alpha=0.1$ by default); Tolerance level of violation rate $\delta$ ($\delta = 0.1$ by default)
        \end{itemize}}
    \Output{The estimated threshold $\hat{C}_\alpha$ for estimated NP classifier $\hat{\phi}_\alpha\left(\cdot\right) = \bbone\left\{ \hat{s}(\cdot) > \hat{C}_\alpha \right\}$.}
    
    \For{$k = n_0''$ to $1$}{
        Calculate violation rate upper bound $v(k) = \sum_{j=k}^{n_0''} \binom{n_0''}{j} (1-\alpha)^j \alpha^{n_0''-j}$ \\
        If $v(k) \leq \delta$ let $k^\ast = k$, break;
    }
    Calculate the set of scores $T = \{t_1, t_2, \ldots, t_{n_0''}\} = \{\hat{s}(\calX) ,\quad \calX \in S_0'' \}$ \\
    Sort $T$ with increasing order, $T = \{t_{(1)}, t_{(2)}, \ldots, t_{(n_0'')}\}$ \\
    The estimated threshold $\hat{C}_\alpha = t_{(k^\ast)}$ 
    \caption{Umbrella Algorithm\citep{tong2018neyman}}
    \label{alg:umbrella}
\end{algorithm}

\section{Additional Numerical Results}
    \subsection{Simulation}
    We provide additional numerical results in this section, including simulation results under imbalanced class sizes, varying tensor shape (Example~\ref{exp:vary shape}), misspecified rank (Example~\ref{exp:vary rank}) and an additional Example~\ref{exp:vary distribution} examining the robustness against non-Gaussian distribution.

\begin{figure}[htbp]
        \centering
        \includegraphics[width=1\textwidth]{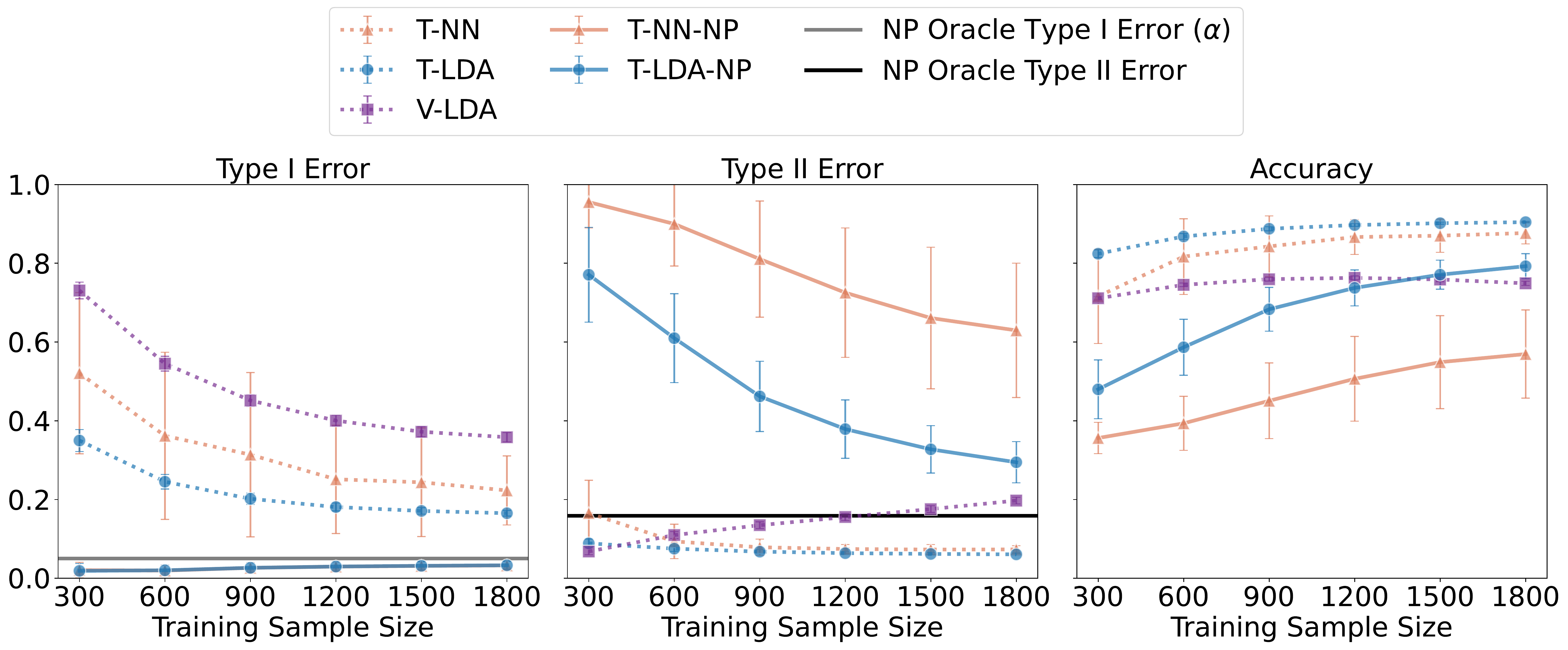}
        \caption{Average Type I Error, Average Type II Error and Accuracy under class size ratio $\eta=2$ ($\alpha=0.05, \delta=0.1$), comparing methods with increasing size of training data}
        \label{fig:vary train size imbalanced}
    \end{figure}

    We first revisit Example~\ref{exp:vary train size} revisited, with imbalanced class sizes. In FIG~\ref{fig:vary train size imbalanced}, we present the average Type I Error, average Type II Error and accuracy under class size ratio $\eta=2$ and all other settings remain the same as in Example~\ref{exp:vary train size}. Although the type I errors of non-NP methods are exerberated under imbalanced class sizes, our proposed NP methods still control the type I errors under the pre-specified level $\alpha=0.05$ with violation rates below $\delta=0.1$ as shown in Table \ref{tbl:vary train size imbalanced}.

    \begin{table}[h!]
    \caption{Violation Rate in Example \ref{exp:vary train size}, ($\alpha=0.05, \delta=0.1$), comparing methods with increasing size of training data under class size ratio $\eta=2$.}
    \label{tbl:vary train size imbalanced}
    \centering
    \begin{tabular}{|c||c|c|c|c|c|}
    \hline
    $n^{\text{train}}$ & T-NN & T-LDA & V-LDA & T-NN-NP & T-LDA-NP\\
    \hline
    300 & 1.000 & 1.000 & 1.000 & .080 & .064\\
    \hline
    600 & 1.000 & 1.000 & 1.000 & .034 & .034\\
    \hline
    900 & 1.000 & 1.000 & 1.000 & .060 & .048\\
    \hline
    1200 & 1.000 & 1.000 & 1.000 & .062 & .074\\
    \hline
    1500 & 1.000 & 1.000 & 1.000 & .052 & .056\\
    \hline
    1800 & 1.000 & 1.000 & 1.000 & .056 & .050\\
    \hline
    \end{tabular}
    \end{table}
    
\begin{figure}[htbp]
        \centering
        \includegraphics[width=1\textwidth]{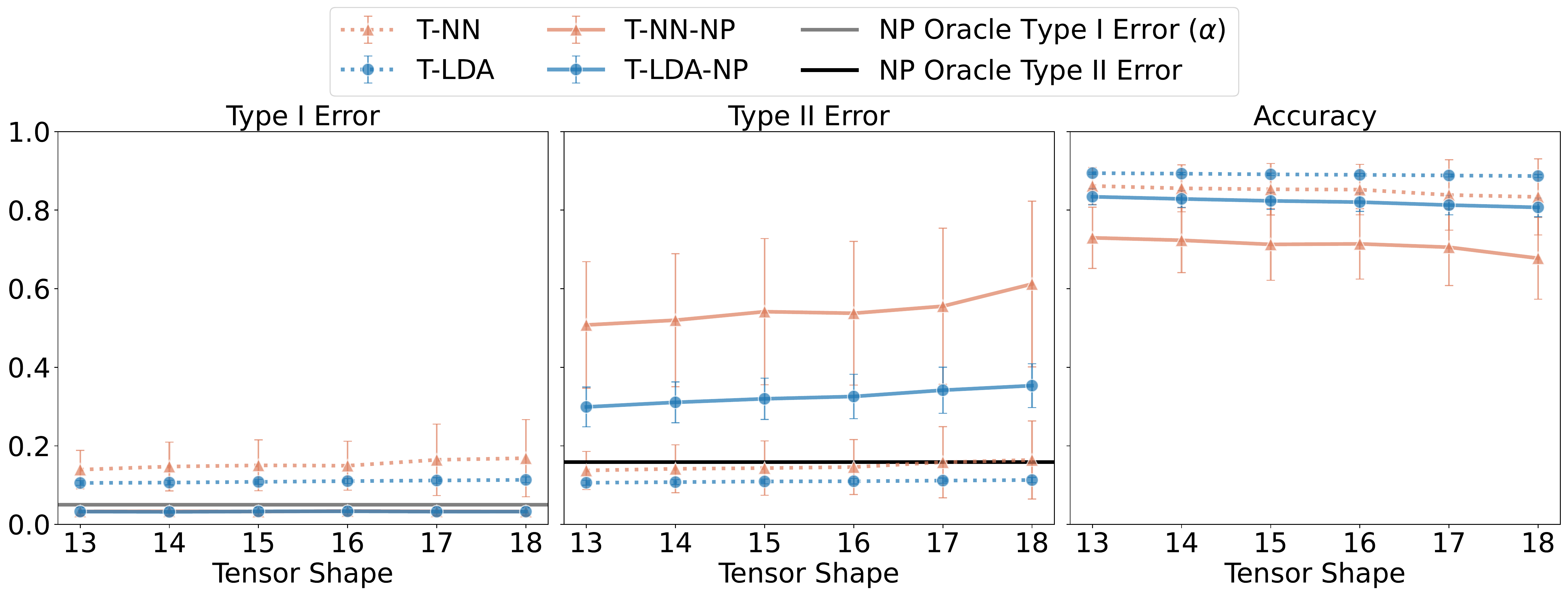}
        \caption{Average Type I Error, Average Type II Error and Accuracy in Example \ref{exp:vary shape}  ($\alpha=0.05, \delta=0,1$), compare methods with increasing $d$. Here $d$ is the scale of tensor shape and the tensor data are in shape of $(d,d,d)$.
        }
        \label{fig:vary shape}
    \end{figure}

    FIG~\ref{fig:vary shape} and Table~\ref{tbl:vary shape} present the results for Example~\ref{exp:vary shape} where we vary the tensor shape. The low-rank structure enable our methods to  maintain robust performance as the tensor dimension increases, with violation rates under $\delta=0.1$ for both the NP classifiers.

    \begin{table}[htbp]
    \caption{Violation Rate in Example \ref{exp:vary shape} ($\alpha=0.05, \delta=0.1$), compare methods with increasing $d$. Here $d$ is the scale of tensor shape and the tensor data are in shape of $(d,d,d)$.}
    \label{tbl:vary shape}
    \centering
    \begin{tabular}{|c||c|c|c|c|}
    \hline
    shape & T-NN & T-LDA & T-NN-NP & T-LDA-NP\\
    \hline
    13 & 1.000 & 1.000 & .070 & .058\\
    \hline
    14 & 1.000 & 1.000 & .082 & .044\\
    \hline
    15 & 1.000 & 1.000 & .064 & .060\\
    \hline
    16 & 1.000 & 1.000 & .078 & .076\\
    \hline
    17 & 1.000 & 1.000 & .088 & .060\\
    \hline
    18 & 1.000 & 1.000 & .062 & .050\\
    \hline
    \end{tabular}
    \end{table}

    Although the T-LDA and T-LDA-NP methods require the specification of the Tucker rank, Example~\ref{exp:vary rank} investigates the robustness of these methods to misspecification of the Tucker rank. Tables~\ref{tbl:varyrank_accuracy}-\ref{tbl:varyrank_violation_rate} summarize the average accuracies, type~I errors, type~II errors and violation rates under different misspecified ranks of T-LDA and T-LDA-NP (as T-NN based method do not require an input rank). In the tables, each row represents a +2 or -2 change in the rank of a particular coordinate, while the ranks of the other coordinates remain unchanged. For example, in the row corresponding to coordinate x, +2 corresponds to the rank vector $(6,\, 6,\, 3)$, and -2 corresponds to the rank vector $(2,\, 6,\, 3)$.
    
    Across all considered working Tucker ranks, both T-LDA and T-LDA-NP demonstrate remarkable robustness to rank misspecification. As expected, the true rank $(4,6,3)$ yields optimal performance, with T-LDA achieving an accuracy of $0.891$. Under misspecification, performance degrades only marginally: T-LDA accuracy ranges from $0.877$ to $0.890$, while T-LDA-NP accuracy ranges from $0.805$ to $0.825$. Crucially, type~I error control remains effective across all rank specifications, with empirical violation rates for T-LDA-NP ranging from $0.044$ to $0.062$, all well below the target threshold $\delta=0.1$. Since the Tucker rank does not affect T-NN or T-NN-NP, we report results only for the LDA-based methods.

    \begin{table}[h!]
        \centering
        \caption{Accuracy with Misspecified Rank ($\alpha=0.05, \delta=0.10$, SNR=$7.0$, True Rank=$(4,\, 6,\, 3)$).
    }
        \label{tbl:varyrank_accuracy}
        \begin{tabular}{|c||c|c||c|c||c|c|}
        \hline
        \multirow{2}{*}{Coordinate} & \multicolumn{2}{c||}{+2} & \multicolumn{2}{c||}{-2} & \multicolumn{2}{c|}{True Rank} \\
        \cline{2-7}
        & T-LDA & T-LDA-NP & T-LDA & T-LDA-NP & T-LDA & T-LDA-NP \\
        \hline
        x & .886 (.003) & .812 (.023) & .884 (.004) & .821 (.021) & \multirow{3}{*}{.891 (.003)} & \multirow{3}{*}{.824 (.022)} \\
        \cline{1-5}
        y & .890 (.003) & .820 (.023) & .890 (.003) & .825 (.021) & & \\
        \cline{1-5}
        z & .883 (.003) & .805 (.022) & .877 (.005) & .812 (.024) & & \\
        \hline
        \end{tabular}
    \end{table}

    \begin{table}[h!]
        \centering

        \caption{Type I Error with Misspecified Rank ($\alpha=0.05, \delta=0.10$, SNR=$7.0$, True Rank=$(4,\, 6,\, 3)$).
        }
        \label{tbl:varyrank_type_i_error}

        \begin{tabular}{|c||c|c||c|c||c|c|}
        \hline
        \multirow{2}{*}{Coordinate} & \multicolumn{2}{c||}{+2} & \multicolumn{2}{c||}{-2} & \multicolumn{2}{c|}{True Rank} \\
        \cline{2-7}
        & T-LDA & T-LDA-NP & T-LDA & T-LDA-NP & T-LDA & T-LDA-NP \\
        \hline
        x & .113 (.006) & .033 (.010) & .115 (.007) & .033 (.010) & \multirow{3}{*}{.108 (.006)} & \multirow{3}{*}{.033 (.010)} \\
        \cline{1-5}
        y & .110 (.006) & .033 (.010) & .109 (.006) & .033 (.010) & & \\
        \cline{1-5}
        z & .117 (.006) & .033 (.010) & .122 (.007) & .033 (.010) & & \\
        \hline
        \end{tabular}
    \end{table}

    \begin{table}[h!]
        \centering

        \caption{Type II Error with Misspecified Rank ($\alpha=0.05, \delta=0.10$, SNR=$7.0$, True Rank=$(4,\, 6,\, 3)$).
        }
        \label{tbl:varyrank_type_ii_error}

        \begin{tabular}{|c||c|c||c|c||c|c|}
        \hline
        \multirow{2}{*}{Coordinate} & \multicolumn{2}{c||}{+2} & \multicolumn{2}{c||}{-2} & \multicolumn{2}{c|}{True Rank} \\
        \cline{2-7}
        & T-LDA & T-LDA-NP & T-LDA & T-LDA-NP & T-LDA & T-LDA-NP \\
        \hline
        x & .114 (.006) & .344 (.054) & .116 (.007) & .326 (.051) & \multirow{3}{*}{.109 (.006)} & \multirow{3}{*}{.320 (.052)} \\
        \cline{1-5}
        y & .111 (.006) & .326 (.055) & .110 (.006) & .317 (.050) & & \\
        \cline{1-5}
        z & .118 (.007) & .357 (.052) & .123 (.008) & .344 (.057) & & \\
        \hline
        \end{tabular}
    \end{table}

    \begin{table}[h!]
        \centering

        \caption{Violation Rate with Misspecified Rank ($\alpha=0.05, \delta=0.10$, SNR=$7.0$, True Rank=$(4,\, 6,\, 3)$).
        }
        \label{tbl:varyrank_violation_rate}

        \begin{tabular}{|c||c|c||c|c||c|c|}
        \hline
        \multirow{2}{*}{Coordinate} & \multicolumn{2}{c||}{+2} & \multicolumn{2}{c||}{-2} & \multicolumn{2}{c|}{True Rank} \\
        \cline{2-7}
        & T-LDA & T-LDA-NP & T-LDA & T-LDA-NP & T-LDA & T-LDA-NP \\
        \hline
        x & 1.000 & .062 & 1.000 & .056 & \multirow{3}{*}{1.000} & \multirow{3}{*}{.060} \\
        \cline{1-5}
        y & 1.000 & .050 & 1.000 & .044 & & \\
        \cline{1-5}
        z & 1.000 & .056 & 1.000 & .058 & & \\
        \hline
        \end{tabular}
    \end{table}

\begin{figure}[htbp]
        \centering
        \includegraphics[width=1\textwidth]{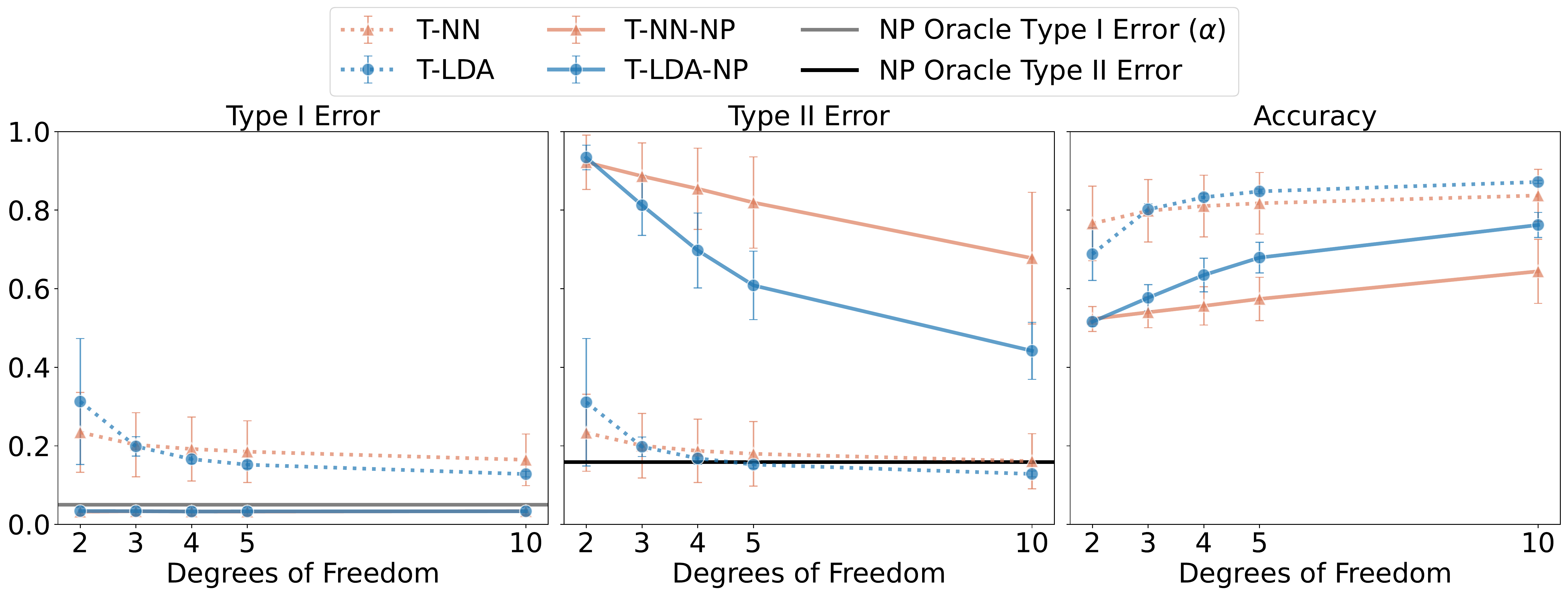}
        \caption{Average Type I Error, Average Type II Error and Accuracy in Example \ref{exp:vary distribution}  ($\alpha=0.05, \delta=0,1$), compare methods with increasing $f$. Here $f$ is the degree of freedom of tensor t-distribution (tensorized multivariate t-distribution). The example aims to examine the robustness of the competing methods when the true underlying distribution deviates from the assumption made in T-LDA.
        }
        \label{fig:vary distribution}
    \end{figure}

    \begin{example}
        \label{exp:vary distribution}
        In this experiment, we examine the robustness of competing methods when the true distribution deviates from T-LDA's Gaussian assumption. Data are generated from a 2-class tensor t-distribution (tensorized multivariate t-distribution) with common $\bSigma$. The tensor shape is $(15,15,15)$ and the degree of freedom is denoted as $f$ and the mean tensor and covariance matrices follow the same scheme as in previous experiments. We vary $f \in \{2,3,4,5,10\}$. The average type I error, average type II error and accuracy are plotted in FIG \ref{fig:vary distribution} and the violation rate is summarized in Table \ref{tbl:vary distribution}.
    \end{example}

    \begin{table}[h!]

    \caption{Violation Rate in Example \ref{exp:vary distribution} ($\alpha=0.05, \delta=0.10$), compare methods with increasing $f$. Here $f$ is the degree of freedom of tensor t-distribution (tensorized multivariate t-distribution). }
    \label{tbl:vary distribution}
    \centering
    \begin{tabular}{|c||c|c|c|c|}
    \hline
    f & T-NN & T-LDA & T-NN-NP & T-LDA-NP\\
    \hline
    2 & 1.000 & .990 & .068 & .084\\
    \hline
    3 & 1.000 & 1.000 & .064 & .076\\
    \hline
    4 & 1.000 & 1.000 & .056 & .076\\
    \hline
    5 & 1.000 & 1.000 & .066 & .068\\
    \hline
    10 & 1.000 & 1.000 & .074 & .060\\
    \hline
    \end{tabular}

    \end{table}

    Similarly as in other examples, both T-NN-NP and T-LDA-NP demonstrate average type I errors below $0.05$ (i.e. $\alpha$), with all violation rates of the NP methods remaining below $0.1$ (i.e. $\delta$). Another noteworthy observation is that the performance of T-LDA and T-LDA-NP improves as the true underlying distribution becomes more similar to the assumed TGMM.

    \subsection{Real Data: Chemical Carcinogenicity (PTC\_MM)}
    The PTC\_MM dataset was collected from the Predictive Toxicology Challenge (PTC) 2000--2001 \citep{helma2001predictive} and contains chemical compounds classified by their observed carcinogenicity in male mice (MM). In this study, we adopt the data provided in \cite{Morris+2020}. To reflect the practical priority of mitigating potential risks to human health, we label compounds with \emph{positive} results in the rodent carcinogenicity test as class~0. This yields $n_0 = 129$ carcinogenic compounds (class~0), and $n_1 = 207$ non-carcinogenic compounds (class~1).
    
    \begin{table}[h!]
    \centering
    \caption{Results on the PTC\_MM dataset.}
    \label{tb:PTC_MM}
    \begin{tabular}{|c||c|c|c|c|c|c|}
    \hline
    & \multicolumn{2}{c|}{$\alpha = 0.40$, $\delta = 0.30$} & \multicolumn{2}{c|}{$\alpha = 0.40$, $\delta = 0.40$} & \multicolumn{2}{c|}{$\alpha = 0.40$, $\delta = 0.50$} \\
    \hline
    Algorithm & T-NN & T-NN-NP & T-NN & T-NN-NP & T-NN & T-NN-NP \\
    \hline
    Type II Error & .112(.074) & .440(.118) & .112(.074) & .422(.130) & .112(.074) & .393(.107) \\
    \hline
    Accuracy & .704(.037) & .604(.066) & .704(.037) & .609(.064) & .704(.037) & .616(.066) \\
    \hline
    Violation Rate & .940 & .240 & .940 & .250 & .940 & .300 \\
    \hline
    \end{tabular}
    \end{table}
    
    Table~\ref{tb:PTC_MM} reports the type~II error, overall accuracy, and the type~I error violation rates across three $(\alpha,\delta)$ configurations with fixed $\alpha = 0.40$ to balance the accuracy in classification of carcinogenic compounds while managing false positives. The standard T-NN approach yields a violation rate of $0.94$ across all settings, demonstrating that type~I error control fails substantially at the nominal level $\alpha$. With T-NN-NP, empirical violation rates drop to between $0.24$ and $0.30$, consistently staying below their respective target thresholds $\delta$ and confirming effective type~I error control. The corresponding accuracy of T-NN-NP ranges from $0.604$ to $0.616$, representing a moderate loss relative to T-NN's $0.704$ accuracy. This trade-off reflects the practical cost of achieving reliable type~I error control in toxicological screening, where misclassifying carcinogenic compounds as safe (false negatives) poses serious health risks. As $\delta$ increases from $0.30$ to $0.50$, we observe a gradual improvement in type~II error (from $0.440$ to $0.393$) and accuracy (from $0.604$ to $0.616$), suggesting that the more relaxed violation tolerance allows the NP method to adopt a less conservative threshold. 

\section{Proofs} \label{sec-proof}
For simplicity, all $\hat{s}$, $\hat{\phi}$ and $\hat{C}_\alpha$ refer to  $\hat{s}^{\text{ LDA-NP}}$, $\hat{\phi}_\alpha^{\text{ LDA-NP}}$ and $\hat{C}_\alpha^{\text{ LDA-NP}}$, respectively.
    \subsection{Oracle Scoring Function, Lemma~\ref{lem:oracle score function}}
        \begin{proof}
        From $\vect(\calX) \sim \cN(\vect(\cM_0); \bSigma_v)$, $C_\alpha^{\ast\ast}$ could be computed: 
        \begin{equation*}
        \begin{aligned}
        s^\ast (\calX) &= \langle \calX , \; \calB \rangle 
        = \vect(\calB)^\top \vect(\calX) \\
        & \sim \cN\left(\vect(\calB)^\top \vect(\cM_0); \vect(\calB)^\top \left(\Sigma_M\otimes\cdots\otimes\Sigma_1 \right) \vect(\calB) \right).
        \end{aligned} 
        \end{equation*}
        We could plugging in $\calB = \cD \times_{m=1}^M \Sigma_m^{-1}$:

        \begin{equation}\label{eqn:vectorized B}
        \begin{aligned}
        \vect(\calB) &= \vect\left(\text{mat}_1(\calB)\right) \\
        &= \vect\left(\Sigma_1^{-1} \text{mat}_1(\cD) \left( \Sigma_M^{-1} \otimes \ldots \otimes \Sigma_2^{-1}\right) ^\top \right) \\
        &= \left( \Sigma_M^{-1} \otimes \ldots \otimes \Sigma_2^{-1} \otimes I_{d_1}\right) \vect\left(\Sigma_1^{-1} \text{mat}_1(\cD) \right)\\
        &= \left( \Sigma_M^{-1} \otimes \ldots \otimes \Sigma_2^{-1} \otimes I_{d_1}\right) \left(I_{d_{-1}} \otimes \Sigma_1^{-1}\right)\vect\left(\text{mat}_1(\cD) \right)\\
        &= \bSigma_v^{-1} \vect\left(\cD \right).
        \end{aligned}
        \end{equation}
        And the Covariance matrix could be simplified:

        \begin{equation*}
        \begin{aligned}
        &\vect(\calB)^\top \left(\Sigma_M\otimes\cdots\otimes\Sigma_1 \right) \vect(\calB) \\
        =& \vect\left(\cD \right)^\top \bSigma_v^{-1} \bSigma_v \bSigma_v^{-1} \vect\left(\cD \right) \\
        =& \vect\left(\cD \right)^\top  \bSigma_v^{-1} \vect\left(\cD \right). 
        \end{aligned}
        \end{equation*}
        In conclusion, given class $0$

        \begin{equation}\label{eqn:normal distribution of oracle score funcion}
        \begin{aligned}
        s^\ast (\calX) \sim \cN\left(\vect(\calB)^\top \vect(\cM_0); \vect\left(\cD \right)^\top  \bSigma_v^{-1} \vect\left(\cD \right)  \right),
        \end{aligned} 
        \end{equation}
        and the threshold $C_\alpha^{\ast\ast}$ 

        \begin{equation}\label{eqn:oracle threshold}
        \begin{aligned}
        C_\alpha^{\ast\ast} = \sqrt{\vect\left(\cD \right)^\top \bSigma_v^{-1} \vect\left(\cD \right)  } \Phi^{-1} (1-\alpha) + \vect(\calB)^\top \vect(\cM_0) \text{,}\\
        \text{where $\Phi (\cdot)$ is the CDF of standard normal distribution}
        \end{aligned} 
        \end{equation}
        so that we have $\Pr\left(\phi _\alpha^\ast (\calX)=1|Y=0\right) = \Pr\left(s^\ast (\calX)> C_\alpha^{\ast\ast}|Y=0\right) = \alpha$. 
        \end{proof}

    In following proofs, we particularly denote the upper bound of $\|\calB\|_F$ and the lower bound of the signal-to-noise ratio, derived from Assumption~\ref{aspt:bounded norm and snr}, as $\|\calB\|_F \leq C_B$ and $\vect(\cD)^\top (\bSigma_v^{-1} ) \vect(\cD) \geq C_L$.
    
    \subsection{Tucker Estimation of Discriminant Tensor $\calB$, Proposition~\ref{theorem: tucker}}
    \begin{proof}
        The proposition is analogous to Theorem 4.1 in \cite{chen2024higha}, which was established for the missing data case, and thus can be derived in a similar manner.
    \end{proof}
        
    \subsection{High Probability Upper Bound for Deviation of Scoring Function,  Theorem~\ref{thm:deviation of s}}
        \begin{proof}
            Consider the decomposition:
            \begin{equation*}
                \begin{aligned}
                    \| s^\ast - \hat{s} \|_{\infty, \mathcal{C}} =& \max_{\calX\in\mathcal{C}} |s^\ast(\calX) - \hat{s}(\calX)|
            = \max_{\calX\in\mathcal{C}} | \langle \calX,\calB \rangle - \langle \calX, \hat{\calB\rangle}| \\
            \leq& \max_{\calX\in\mathcal{C}} | \langle \calX-\calM,\calB-\hat{\calB\rangle}| + |\langle \calM,\calB-\hat{\calB\rangle} | .
                \end{aligned}
            \end{equation*}   
            The second term does not contain $\calX$ and hence is irrelevant to $\mathcal{C}$. Notice that for $\calX$ from the class $i$, $i\in\{0,1\}$, \[\angles{\calX-\calM, \hat{\calB}-\calB} \sim \cN(\rm vec(\hat{\calB}-\calB)^\top \rm vec(\cM_i-\cM), \rm vec(\hat{\calB}-\calB)^\top \bSigma_v \rm vec(\hat{\calB}-\calB)) .\]
            We know that
            $\angles{\calX-\calM, \hat{\calB}-\calB}$ is $\sqrt{\norm{\bSigma_v}}\norm{\hat{\calB}-\calB}_F$-sub-Gaussian. 
            By assumption, $\norm{\bSigma_v}\le\prod_{m\in[M]}\norm{\bSigma_m}\le C_0^M$ is bounded.
            Thus, applying a sub-Gaussian inequality, we get that 
            \begin{equation*}
            \Pr_i\paran{\left| \angles{\calX-\calM, \hat{\calB}-\calB} \right|/(\sqrt{\norm{\bSigma_v}}\norm{\hat{\calB}-\calB}_F) \ge t} \le 2 \exp(-t^2/2)\quad\text{for }i=0,1.
            \end{equation*}
            Under the assumptions in Theorem~\ref{thm:deviation of s}, by Proposition~\ref{theorem: tucker}, we have $\| \hat{\calB}-\calB \|_F \asymp \sqrt{\frac{d_0 r_0}{n_{\text{min}}}}$. Hence
            \[
                \Pr_i(\calX \notin \mathcal{C}) \leq  2 \exp(-t^2/2)\asymp \exp(-(n_0\wedge n_1)^{(2a)})\quad\text{for }i=0,1,
            \]
            and
            \[ \| s^\ast - \hat{s} \|_{\infty, \mathcal{C}}  \leq t\|\hat{\calB}-\calB\|_F+\|\hat{\calB}-\calB\|_F \lesssim t\sqrt{\frac{d_0 r_0}{n_{\text{min}}}} .\]
            Then we could define the high probability set $\mathcal{C}$:
            \begin{equation}\label{eqn:high probabilty set C}
                \mathcal{C} = \left\{\left| \angles{\calX-\calM, \hat{\calB}-\calB}\right|/(\sqrt{\norm{\bSigma_v}}\norm{\hat{\calB}-\calB}_F) < t \right\},
            \end{equation}
            and
            \begin{equation}\label{iqn: probabilty of set C}
                \Pr(\mathcal{C}) > 1-2\exp(-t^2/2).
            \end{equation}
        \end{proof}

    \subsection{Conditional Margin Assumption, Proposition~\ref{prop:margin assumption}}
        \begin{proof}
        \begin{equation*}
            \begin{aligned}
                &\Pr \left\{\left|s^*(\calX)-C_\alpha^{**}\right| \leq \iota \mid \calX \in \mathcal{C}\right\} \\
                \leq& \Pr \left\{\left|s^*(\calX)-C_\alpha^{**}\right| \leq \iota \right\}/\Pr(\calX \in \mathcal{C})\\
                =& (\Phi(U)-\Phi(L))/\Pr(\calX \in \mathcal{C}),
            \end{aligned}
        \end{equation*}
        where $\Phi$ is the cumulative distribution function of standard norm distribution, $U = (C_\alpha^{**}+\iota-\vect(\calB)^\top \vect(\cM_0))/\sqrt{\vect\left(\cD \right)^\top \bSigma_v^{-1} \vect\left(\cD \right)}$, and\\$L = (C_\alpha^{**}-\iota-\vect(\calB)^\top \vect(\cM_0))/\sqrt{\vect\left(\cD \right)^\top \bSigma_v^{-1} \vect\left(\cD \right)}$. The last equation utilized \eqref{eqn:normal distribution of oracle score funcion}. By the Mean Value Theorem, 
        \[\Phi(U)-\Phi(L) = \phi(z)(U-L) = \phi(z)\frac{2\iota}{\sqrt{\vect\left(\cD \right)^\top \bSigma_v^{-1} \vect\left(\cD \right)}} ,\]
        where $\phi$ is the probability density function standard norm distribution and $L \leq z \leq U$, thus,
        \begin{equation}
            \label{iqn: margin assumption}
            \Pr \left\{\left|s^*(\calX)-C_\alpha^{**}\right| \leq \iota \mid \calX \in \mathcal{C}\right\}
            \leq \frac{2\phi(0)}{\Pr(\calX \in \mathcal{C})\sqrt{C_L}} \iota\leq \frac{4\phi(0)}{\sqrt{C_L}}\iota,
        \end{equation}
        where $C_L$ denotes the lower bound of the signal-to-noise ratio and the last inequality holds as $\Pr(\calX \in \mathcal{C}) \geq 1/2$.
        \end{proof}

    \subsection{Conditional Detection Condition, Proposition~\ref{prop:conditional detection condition}}
        \begin{proof}
    We first define events $\mathcal{E}$ as following:
    \begin{equation*}
        \begin{aligned}
            \mathcal{E} =& \{ \calX: C_\alpha^{\ast\ast} \leq s^*(\calX) \leq C_\alpha^{\ast\ast}+\iota \}.
        \end{aligned}
    \end{equation*}
    Then, 
    \begin{equation*}
        \begin{aligned}
            \Pr \left( \mathcal{E} \right) 
            = \Phi(U')-\Phi(L'),
        \end{aligned}
    \end{equation*}
    where $\Phi$ is the cummulative distribution function of standard norm distribution, 
    \begin{equation*}
        \begin{aligned}
            U' =& (C_\alpha^{**}+\iota-\vect(\calB)^\top \vect(\cM_0))/\sqrt{\vect\left(\cD \right)^\top \bSigma_v^{-1} \vect\left(\cD \right)} \\
            =& \Phi^{-1} (1-\alpha)+\frac{\iota}{\sqrt{\vect\left(\cD \right)^\top \bSigma_v^{-1} \vect\left(\cD \right)}},
        \end{aligned}
    \end{equation*}
    
    and, \[L' = (C_\alpha^{**}-\vect(\calB)^\top \vect(\cM_0))/\sqrt{\vect\left(\cD \right)^\top \bSigma_v^{-1} \vect\left(\cD \right)}=\Phi^{-1} (1-\alpha).\]The last equation utilized \eqref{eqn:normal distribution of oracle score funcion} and the calculation of $U'$ and $L'$ utilized \eqref{eqn:oracle threshold}. By the Mean Value Theorem, 
    \[\Phi(U')-\Phi(L') = \phi(z')(U'-L') = \phi(z')\frac{\iota}{\sqrt{\vect\left(\cD \right)^\top \bSigma_v^{-1} \vect\left(\cD \right)}} ,\]
    where $\phi$ is the probabiltiy density function standard norm distribution and $L' \leq z' \leq U'$, notice that normally $\alpha<1/2$,
    \begin{equation}
        \label{iqn: detection condition-Upper}
        \Pr \left( \mathcal{E} \right)
        \geq \frac{\phi(\Phi^{-1} (1-\alpha)+\frac{\delta^*}{C_L})}{\sqrt{C_U}}\iota ,
    \end{equation}
    $C_U$ is the upper bound of $\vect\left(\cD \right)^\top  \bSigma_v^{-1}  \vect\left(\cD \right)$ and it could be given by $C_B$, the upper bound of $\|\calB\|_F$, and $C_0^M$, the upper bound of $\bSigma_v$: $\vect\left(\cD \right)^\top  \bSigma_v^{-1}  \vect\left(\cD \right) = \vect\left(\calB \right)^\top  \bSigma_v  \vect\left(\calB \right) \leq C_U \stackrel{\text{def}}{=} C_B^2C_0^M$. 
    Then by
    \begin{equation}\label{eqn:detection to conditional detection}
        \begin{aligned}
            &\Pr \left( \mathcal{E} \mid \mathcal{C}\right) 
            = \frac{\Pr \left( \mathcal{C} \mid \mathcal{E}\right)}{\Pr(\mathcal{C})}\Pr(\mathcal{E}),
        \end{aligned}
    \end{equation}
    it is sufficient to show that $\Pr(\mathcal{C}\mid\mathcal{E})$ is bounded by below, and immediately,
    \begin{equation*}
        \Pr \left( \mathcal{E} \mid \mathcal{C}\right)
            \geq C \Pr(\mathcal{E}),
    \end{equation*}
    where C is some constant.\\
    To obtain such lower bound, we first find a restricted set $\mathcal{C}'$ satisfying:
    \begin{equation} \label{requirement of C'}
        \begin{aligned}
            \mathcal{C}' \cap \mathcal{E} &\subset \mathcal{C} \cap \mathcal{E},\\
            \mathcal{C}' &\perp \!\!\!\! \perp \mathcal{E}. 
        \end{aligned}
    \end{equation}
    $\mathcal{C}'$ could be constructed by considering the following random variable:
    \begin{equation*}
        \begin{aligned}
            \langle \calX, \calB' \rangle = \vect(\calB')^\top \vect(\calX),
        \end{aligned}
    \end{equation*}
    where $\calB' = \lambda \calB + (1-\lambda)\hat{\calB}$ and $\lambda = \frac{\langle \cD, \hat{\calB}\rangle}{\langle \cD, \hat{\calB}-\calB \rangle}$. We will first show that $\langle \calX, \calB' \rangle$ and $\langle \calX, \calB \rangle$ are independent, then we will rule out the cases which entail $\langle \cD, \hat{\calB}-\calB \rangle = 0$, to ensure that $\lambda$ is well-defined. Finally, we will construct $\mathcal{C}'$ via $\langle \calX, \calB' \rangle$.\\
    Notice that, $\langle \calX, \calB' \rangle$ and $\langle \calX, \calB \rangle$ follow jointly normal distribution:
    \begin{equation*}
    \begin{aligned}
            \begin{bmatrix}
                \langle \calX, \calB \rangle \\
                \langle \calX, \calB' \rangle
            \end{bmatrix}&=
            \begin{bmatrix}
                \vect(\calB)^\top \\
                \vect(\calB')^\top
            \end{bmatrix} \vect(\calX) \\
            &\sim \mathcal{N}\left( 
                \begin{bmatrix}
                    \vect(\calB)^\top \\
                    \vect(\calB')^\top
                \end{bmatrix} \vect(\cM_0), \; 
                \begin{bmatrix}
                    \vect(\calB)^\top \\
                    \vect(\calB')^\top
                \end{bmatrix} \bSigma_v
                \begin{bmatrix}
                    \vect(\calB) &
                    \vect(\calB')
                \end{bmatrix}
            \right).
        \end{aligned}
    \end{equation*}
    And the covariance term:
    \begin{equation*}
        \begin{aligned}
            \vect(\calB)^\top \Sigma_v \vect(\calB')
            =& \left( \bSigma_v \vect(\calB) \right)^\top \vect(\calB')
            \stackrel{\eqref{eqn:vectorized B}}{=} \vect(D)^\top \vect(\calB')\\
            =&\frac{\langle \cD, \hat{\calB}\rangle}{\langle \cD, \hat{\calB}-\calB \rangle} \langle \cD, \calB \rangle -
            \frac{\langle \cD, \calB\rangle}{\langle \cD, \hat{\calB}-\calB \rangle}  \langle \cD, \hat{\calB} \rangle
            =0.
        \end{aligned}
    \end{equation*}
    Hence $\langle \calX, \calB' \rangle$ and $\langle \calX, \calB \rangle$ are independent. Next we discuss the cases that $\langle \cD, \hat{\calB}-\calB \rangle = 0$.\\
    
    \textbf{Case 1}: $\hat{\calB} = c \calB$ for any constant $c \neq 0$. (Ruling out this case would ensure $\hat{\calB}-\calB \neq 0$, furthermore, $\calB'\neq 0$)

    $\mathcal{C}$ would reduced to:
    \[\mathcal{C} = \left\{\left| \angles{\calX-\calM, \calB}\right|/(\sqrt{\norm{\bSigma_v}}\norm{\calB}_F) < t \right\}.\]
    Combined with \eqref{eqn:oracle threshold}, we have $\mathcal{E}\subset \mathcal{C}$ as long as we choose $\delta^* \in (0, t(\sqrt{\norm{\bSigma_v}}\norm{\calB}_F)-|\langle \cM,\calB \rangle-C_\alpha^{**}|)$ and the further assumption $t \asymp (n_0\wedge n_1)^a > \Phi^{-1}(1-\alpha)$ would ensure the feasibility,
    \[\Pr \left(\mathcal{E} \mid \mathcal{C}\right) 
            = \frac{\Pr \left( \mathcal{C} \cap \mathcal{E}\right)}{\Pr(\mathcal{C})} = \frac{\Pr \left( \mathcal{E}\right)}{\Pr(\mathcal{C})} \geq \Pr \left( \mathcal{E}\right).\] 
    Thus, the desired \eqref{eqn:detection to conditional detection} holds.\\
    
    \textbf{Case 2}: $\hat{\calB}-\calB \neq 0, \langle \cD, \hat{\calB}-\calB \rangle = \vect(\cD)^\top \vect(\hat{\calB}-\calB)=0$.
    
    By the same argument regarding the jointly normal distribution, $\langle \calX, \hat{\calB}-\calB \rangle$ and $\langle \calX, \calB \rangle$ are independent, then $\mathcal{C} \perp \!\!\!\! \perp \mathcal{E}$. $\mathcal{C}' := \mathcal{C}$ in this case.\\
    
    Now we could continue to construct $\mathcal{C}'$,
    \[\mathcal{C}' = \left\{ \calX: L'' \leq \langle \calX-\cM_0, \calB' \rangle \leq U'' \right\}.\]
    Since we have already shown that $\langle \calX, \calB' \rangle$ and $\langle \calX, \calB \rangle$ are independent, $\mathcal{C}'$ and $\mathcal{E}$ are independent, as expected. It remains to 
     determine $L''$ and $U''$, such that $\calX \in \mathcal{C}'\cap \mathcal{E} \Rightarrow \calX \in \mathcal{C}\cap \mathcal{E}$. To this end, we need to decompose $\langle \calX-\cM, \frac{\hat{\calB}-\calB}{\|\hat{\calB}-\calB\|_F} \rangle$ in terms of $\langle\calX-\cM_0, \calB'\rangle$ and $\langle \calX, \calB \rangle$, recall that $\cM = (\cM_0 + \cM_1)/2$, $\cD=(\cM_1-\cM_0)$:
    \begin{equation*}
        \begin{aligned}
            &\left\langle \calX-\cM, \frac{\hat{\calB}-\calB}{\|\hat{\calB}-\calB\|_F} \right\rangle \\
            =&\langle \calX-\cM_0, \frac{\hat{\calB}-\calB}{\|\hat{\calB}-\calB\|_F} \rangle - \langle \cD/2, \frac{\hat{\calB}-\calB}{\|\hat{\calB}-\calB\|_F} \rangle \\
            =& \left\langle \calX-\cM_0, \frac{1}{\|\hat{\calB}-\calB\|_F}\left( \frac{\calB'-\calB}{1-\lambda} \right) \right\rangle - \langle \cD/2, \frac{\hat{\calB}-\calB}{\|\hat{\calB}-\calB\|_F} \rangle \\
            =& \frac{1}{\|\hat{\calB}-\calB\|_F}\frac{1}{1-\lambda}\left\langle \calX-\cM_0, \calB' \right\rangle - \frac{1}{\|\hat{\calB}-\calB\|_F}\frac{1}{1-\lambda}\left\langle \calX-\cM_0, \calB \right\rangle - \langle \cD/2, \frac{\hat{\calB}-\calB}{\|\hat{\calB}-\calB\|_F} \rangle \\
            =& -\frac{1}{\|\hat{\calB}-\calB\|_F}\frac{\langle \cD, \hat{\calB}-\calB \rangle}{\langle\cD, \calB \rangle}\left\langle \calX-\cM_0, \calB' \right\rangle + \frac{1}{\|\hat{\calB}-\calB\|_F}\frac{\langle \cD, \hat{\calB}-\calB \rangle}{\langle\cD, \calB \rangle}\left\langle \calX-\cM_0, \calB \right\rangle \\
            &-\langle \cD/2, \frac{\hat{\calB}-\calB}{\|\hat{\calB}-\calB\|_F} \rangle \\
            =& -\frac{\left\langle \cD, \frac{\hat{\calB}-\calB}{\|\hat{\calB}-\calB\|_F} \right\rangle }{\langle\cD, \calB \rangle}\left\langle \calX-\cM_0, \calB' \right\rangle + \frac{\left\langle \cD, \frac{\hat{\calB}-\calB}{\|\hat{\calB}-\calB\|_F} \right\rangle}{\langle\cD, \calB \rangle}\left\langle \calX-\cM_0, \calB \right\rangle - \langle \cD/2, \frac{\hat{\calB}-\calB}{\|\hat{\calB}-\calB\|_F} \rangle \\
            =& -\rho\left( \langle \calX-\cM_0, \calB' \rangle - \langle \calX-\cM_0, \calB \rangle +\frac{1}{2}\langle\cD,\calB\rangle \right).
        \end{aligned}
    \end{equation*}
    The decomposition is well-defined ($\lambda \neq 1$), since $\lambda = 1$ implies $\langle \calB, \cD \rangle =0$, which violates the Assumption~\ref{aspt:bounded norm and snr}. And in the last equation we introduced the shorthand notation $\rho = \frac{\left\langle \cD, \frac{\hat{\calB}-\calB}{\|\hat{\calB}-\calB\|_F} \right\rangle }{\langle\cD, \calB \rangle}$ for simplicity. Notice that 
    {\small \begin{equation*}
        \begin{aligned}
            \mathcal{E} &= \{\calX: C_\alpha^{\ast\ast} \leq s^*(\calX) \leq C_\alpha^{\ast\ast}+\iota \} \\
            &\stackrel{\eqref{eqn:oracle threshold}}{=}  \{ \calX: \sqrt{\langle\cD,\calB\rangle}\Phi^{-1}(1-\alpha) + \langle \calB, \cM_0 \rangle \leq \langle \calX,\calB \rangle \leq \sqrt{\langle\cD,\calB\rangle}\Phi^{-1}(1-\alpha) + \langle \calB, \cM_0 \rangle +\iota \} \\
            &= \{ \calX: \sqrt{\langle\cD,\calB\rangle}\Phi^{-1}(1-\alpha)  \leq \langle \calX-\cM_0,\calB \rangle \leq \sqrt{\langle\cD,\calB\rangle}\Phi^{-1}(1-\alpha)  +\iota \}.
        \end{aligned}
    \end{equation*}}
    
    We could define:
    \begin{equation*}
        \begin{aligned}
            \mathcal{C}' = \Bigg\{ \calX: -\frac{\sqrt{\|\bSigma_v\|}}{|\rho|}t + \sqrt{\langle\cD,\calB\rangle}\Phi^{-1}(1-\alpha) + \delta^* - \frac{1}{2}\langle\cD,\calB\rangle < \langle \calX-\cM_0, \calB' \rangle \\
            < \frac{\sqrt{\|\bSigma_v\|}}{|\rho|}t + \sqrt{\langle\cD,\calB\rangle}\Phi^{-1}(1-\alpha) - \frac{1}{2}\langle\cD,\calB\rangle\Bigg\}.
        \end{aligned}
    \end{equation*}
    Then we have 
    \begin{equation*}
        \begin{aligned}
            \calX \in \mathcal{C}' \cap \mathcal{E} \Rightarrow \calX \in \mathcal{C},
        \end{aligned}
    \end{equation*}
    which is equivalent to $\mathcal{C}' \cap \mathcal{E} \subset \mathcal{C}$ and naturally $\mathcal{C}' \cap \mathcal{E} \subset \mathcal{C} \cap \mathcal{E}$. 
    
    \eqref{requirement of C'} is satisfied and hence, 
    \begin{equation*}
        \begin{aligned}
            \Pr \left( \mathcal{E} \mid \mathcal{C}\right) 
            = \frac{\Pr \left( \mathcal{C} \mid \mathcal{E}\right)}{\Pr(\mathcal{C})}\Pr(\mathcal{E})
            \geq \Pr \left( \mathcal{C} \mid \mathcal{E}\right)\Pr(\mathcal{E}) 
            \geq & \Pr \left( \mathcal{C}' \mid \mathcal{E}\right)\Pr(\mathcal{E}) \\
            = & \Pr \left( \mathcal{C}'\right)\Pr(\mathcal{E}) 
            = \left( \Phi(U'')-\Phi(L'')\right)\Pr(\mathcal{E}).
        \end{aligned}
    \end{equation*}
    The last equation comes from the fact that $\langle \calX-\cM_0, \calB'\rangle \sim \mathcal{N} \left(0, \vect(\calB')^\top \bSigma_v \vect(\calB') \right)$ and 
    \begin{equation*}
        \begin{aligned}
            L'' &= \frac{-\frac{\sqrt{\|\bSigma_v\|}}{|\rho|}t + \sqrt{\langle\cD,\calB\rangle}\Phi^{-1}(1-\alpha) + \delta^* - \frac{1}{2}\langle\cD,\calB\rangle}{\sqrt{\vect(\calB')^\top \bSigma_v \vect(\calB')}} ,\\
            U'' &= \frac{\frac{\sqrt{\|\bSigma_v\|}}{|\rho|}t + \sqrt{\langle\cD,\calB\rangle}\Phi^{-1}(1-\alpha) - \frac{1}{2}\langle\cD,\calB\rangle}{\sqrt{\vect(\calB')^\top \bSigma_v \vect(\calB')}} .
        \end{aligned}
    \end{equation*}
    Note that 
    \begin{equation*}
        \begin{aligned}
            \calB' = \lambda \calB + (1-\lambda)\hat{\calB} 
            = \frac{\langle \cD, \calB \rangle}{\langle \cD, \hat{\calB}-\calB \rangle}(\hat{\calB}-\calB)+\calB
            = \frac{1}{\rho} \frac{\hat{\calB}-\calB}{\|\hat{\calB}-\calB\|_F}+\calB .
        \end{aligned}
    \end{equation*}
    The covariance term
    \begin{equation*}
        \begin{aligned}
            &\vect(\calB')^\top \bSigma_v \vect(\calB') \\
            =& \frac{1}{\rho^2}\frac{\vect(\hat{\calB}-\calB)^\top \bSigma_v \vect(\hat{\calB}-\calB)}{\|\hat{\calB}-\calB\|_F^2} + \frac{2}{\rho}\frac{\vect(\hat{\calB}-\calB)^\top \bSigma_v \vect(\calB)}{\|\hat{\calB}-\calB\|_F} + \vect(\calB)^\top \bSigma_v \vect(\calB)\\
            =& \frac{1}{\rho^2}\frac{\vect(\hat{\calB}-\calB)^\top \bSigma_v \vect(\hat{\calB}-\calB)}{\|\hat{\calB}-\calB\|_F^2} + 3\langle \cD, \calB \rangle.
         \end{aligned}
    \end{equation*}
    In the last equation we plug-in the definition of $\rho$ and utilize that $\vect(\cD) = \bSigma_v \vect(\calB)$. One could rewrite:
    \begin{equation*}
        \begin{aligned}
            L'' &= \frac{-\sqrt{\|\bSigma_v\|}t + |\rho|\sqrt{\langle\cD,\calB\rangle}\Phi^{-1}(1-\alpha) + |\rho|\delta^* - \frac{|\rho|}{2}\langle\cD,\calB\rangle}{\sqrt{\frac{\vect(\hat{\calB}-\calB)^\top \bSigma_v \vect(\hat{\calB}-\calB)}{\|\hat{\calB}-\calB\|_F^2} + 3\rho^2\langle \cD, \calB \rangle}},\\
            U'' &= \frac{\sqrt{\|\bSigma_v\|}t + |\rho|\sqrt{\langle\cD,\calB\rangle}\Phi^{-1}(1-\alpha) - \frac{|\rho|}{2}\langle\cD,\calB\rangle}{\sqrt{\frac{\vect(\hat{\calB}-\calB)^\top \bSigma_v \vect(\hat{\calB}-\calB)}{\|\hat{\calB}-\calB\|_F^2} + 3\rho^2\langle \cD, \calB \rangle}}.
        \end{aligned}
    \end{equation*}
    Recall that $\rho = \frac{\left\langle \cD, \frac{\hat{\calB}-\calB}{\|\hat{\calB}-\calB\|_F} \right\rangle }{\langle\cD, \calB \rangle}$, we have $0 \leq |\rho| \leq \frac{\|\cD\|_F}{\langle \cD, \calB \rangle}$. Together with $C_0^{-M} \leq \| \bSigma_v \| \leq C_0^M$ (by Assumption~\ref{aspt:bounded norm and snr}), $\|\calB\|_F \leq C_B$ and $\langle\cD, \calB\rangle \geq C_L$, we conclude that a further assumption on $t \asymp n_{\text{min}}^a$:
    \begin{equation*}
           t\geq C_0^M\sqrt{1+3\frac{C_B}{C_L}}\Phi^{-1}(\frac{3}{4})+\frac{C_0^{3M/2}C_B}{C_L}\left(\delta^*+\sqrt{C_L}\Phi^{-1}(1-\alpha)+C_L/2\right),
    \end{equation*}
    will ensure that{\small
    \begin{equation*}
        t\geq \frac{\sqrt{\frac{\vect(\hat{\calB}-\calB)^\top \bSigma_v \vect(\hat{\calB}-\calB)}{\|\hat{\calB}-\calB\|_F^2} + 3\rho^2\langle \cD, \calB \rangle} \Phi^{-1}(\frac{3}{4}) + |\rho|\delta^* +\left| |\rho|\sqrt{\langle\cD\calB\rangle}\Phi^{-1}(1-\alpha)-\frac{|\rho|}{2}\langle\cD\calB\rangle \right|}{\|\bSigma_v\|},
    \end{equation*}}
    which will lead to 
    \begin{equation*}
        \begin{aligned}
            \Pr \left( \mathcal{E} \mid \mathcal{C}\right)
            \geq \left( \Phi(U'')-\Phi(L'')\right)\Pr(\mathcal{E})
            \geq \left( \frac{3}{4}-\frac{1}{4}\right)\Pr(\mathcal{E})
            \stackrel{\eqref{iqn: detection condition-Upper}}{\geq} \frac{\phi(\Phi^{-1} (1-\alpha)+\frac{\delta^*}{C_L})}{2C_BC_0^{M/2}}\iota .
        \end{aligned}
    \end{equation*}
    
    Follow similar analysis,
    \begin{equation*}
        \Pr \left\{ C_\alpha^{\ast\ast} \geq s^*(\calX) \geq C_\alpha^{\ast\ast}-\iota \mid \calX \in \mathcal{C}\right\}
        \geq C \frac{\phi(\Phi^{-1} (1-\alpha))}{2C_BC_0^{M/2}}\iota .
    \end{equation*}
    As a result, if we assume 
    {\small \begin{equation*}
    t\geq \max\left\{\Phi^{-1}(1-\alpha), C_0^M\sqrt{1+3\frac{C_B}{C_L}}\Phi^{-1}(\frac{3}{4})+\frac{C_0^{3M/2}C_B}{C_L}\left(\delta^*+\sqrt{C_L}\Phi^{-1}(1-\alpha)+C_L/2\right)\right\}, \end{equation*}} 
    then
    \begin{equation}
    \begin{aligned}
        \label{iqn: detection condition}
        \Pr \left\{ C_\alpha^{\ast\ast} \leq s^*(\calX) \leq C_\alpha^{\ast\ast}+\iota \mid \calX \in \mathcal{C}\right\} \wedge \Pr \left\{ C_\alpha^{\ast\ast} \geq s^*(\calX) \geq C_\alpha^{\ast\ast}-\iota \mid \calX \in \mathcal{C}\right\}\\
        \geq C\frac{\phi(\Phi^{-1} (1-\alpha)+\frac{\delta^*}{C_L})}{2C_BC_0^{M/2}}\iota.
    \end{aligned}
    \end{equation}
    \end{proof}

    \subsection{Diminishing Excess Type II error, Theorem~\ref{thm:excess type ii error}}
    \begin{proof}
    The control on type I error is guaranteed by Proposition~\ref{prop:umbrella}. Therefore, we only need to show that the excess type II error vanishes with high probability.
    We utilize the decomposition of excess type II error in \cite{zhao2016neyman}: Let $G^\ast=\left\{s^\ast(\calX)<C_\alpha^{\ast\ast}\right\}$ and $\hat{G}=\left\{\hat{s}(\calX)<\hat{C}_\alpha\right\}$, 
     \begin{equation}\label{eqn:excess-typeii-error}
        \begin{aligned}
        {\Pr}_1(\hat{G})-{\Pr}_1\left(G^*\right) 
        = &\int_{\hat{G}} d P_1-\int_{G^*} d P_1=\int_{\hat{G}} r d P_0-\int_{G^*} r d P_0 \\
        = &\int_{\hat{G}}\left(r-C_\alpha^\ast\right) d P_0+C_\alpha^\ast {\Pr}_0(\hat{G})-\int_{G^*}\left(r-C_\alpha^\ast\right) d P_0-C_\alpha^\ast {\Pr}_0\left(G^*\right) \\
        = &\int_{\hat{G} \backslash {G^*}}\left(r-C_\alpha^\ast\right) d P_0-\int_{{G^*} \backslash \hat{G}}\left(r-C_\alpha^\ast \right) d P_0+C_\alpha^\ast\left\{{\Pr}_0(\hat{G})-{\Pr}_0\left(G^*\right)\right\} \\
        = &\int_{\hat{G} \backslash G^*}\left|r-C_\alpha^\ast\right| d P_0+\int_{G^* \backslash \hat{G}}\left|r-C_\alpha^\ast\right| d P_0+C_\alpha^\ast\left\{R_0\left(\phi_\alpha^*\right)-R_0(\hat{\phi}_{\alpha})\right\},
        \end{aligned}
        \end{equation}
    where $P_0$ ($P_1$) is the probability distribution of $\calX$ given that $\calX$ is from class 0 (class 1) respectively. The decomposition \eqref{eqn:excess-typeii-error} transform the excess type II error to a sum of integrals of $\left|r-C_\alpha^\ast\right|$ on two specific sets and the deviation of type I error. Next, we will introduce several technical results that can help us bound the three terms separately. To obtain an upper bound of the deviation of type I error, we quote the next two Lemmas from \cite{tong2020neyman}.

    \begin{lemma}
    \label{lemma:threshold-2016}
    Let $\alpha, \delta_0 \in(0,1)$ and ${n_0''} \geq 4 /\left(\alpha \delta_0\right)$. For any $\delta_0^{\prime} \in(0,1)$, the distance between $R_0\left(\hat{\phi}_{k^{\prime}}\right)$ and $R_0\left(\phi_\alpha^\ast\right)$ can be bounded as
    $$
    \Pr\left\{\left|R_0\left(\hat{\phi}_{k^{\prime}}\right)-R_0\left(\phi_\alpha^\ast\right)\right|>\xi_{\alpha, \delta_0, {n_0''}}\left(\delta_0^{\prime}\right)\right\} \leq \delta_0^{\prime},
    $$
    where
    $$
    \xi_{\alpha, \delta_0, {n_0''}}\left(\delta_0^{\prime}\right)=\sqrt{\frac{k^{\prime}\left({n_0''}+1-k^{\prime}\right)}{\left({n_0''}+2\right)\left({n_0''}+1\right)^2 \delta_0^{\prime}}}+A_{\alpha, \delta_0}\left({n_0''}\right)-(1-\alpha)+\frac{1}{{n_0''}+1},
    $$
    in which $k^{\prime}=\left\lceil\left({n_0''}+1\right) A_{\alpha, \delta_0}\left({n_0''}\right)\right\rceil$ and $A_{\alpha, \delta_0}\left({n_0''}\right)=\frac{1+2 \delta_0\left({n_0''}+2\right)(1-\alpha)+\sqrt{1+4 \delta_0(1-\alpha) \alpha\left({n_0''}+2\right)}}{2\left\{\delta_0\left({n_0''}+2\right)+1\right\}}$. Moreover, if ${n_0''} \geq \max \left(\delta_0^{-2}, \delta_0^{\prime-2}\right)$, we have $\xi_{\alpha, \delta_0, {n_0''}}\left(\delta_0^{\prime}\right) \leq(5 / 2) {n_0''}^{-1 / 4}$.
    \end{lemma}
    
    \begin{lemma}
    \label{lemma:type I error deviation}
    Under the same assumptions as in Lemma \ref{lemma:threshold-2016} , the distance between $R_0\left(\hat{\phi}_{k^\ast}\right)$ and $R_0\left(\phi_\alpha^\ast\right)$ can be bounded as
    $$
    \Pr\left\{\left|R_0\left(\hat{\phi}_{k^\ast}\right)-R_0\left(\phi_\alpha^\ast\right)\right|>\xi_{\alpha, \delta_0, {n_0''}}\left(\delta_0^{\prime}\right)\right\} \leq \delta_0+\delta_0^{\prime}.
    $$
    If ${n_0''} \geq \max \left\{\delta_0^{-2}, \delta_0^{\prime-2},4 /\left(\alpha \delta_0\right)\right\}$, we have $\xi_{\alpha, \delta_0, {n_0''}}\left(\delta_0^{\prime}\right) \leq(5 / 2) {n_0''}^{-1 / 4}$.
    \end{lemma}
     As the tensor data structure as well as the TGMM model fit the assumptions. A high probability upper bound of the deviation of type I error could be directly attained. Hence in \eqref{eqn:excess-typeii-error}
     
        \begin{equation*}
        \begin{aligned}
        {\Pr}_1(\hat{G})-{\Pr}_1\left(G^*\right) 
        = &\int_{\hat{G} \backslash G^*}\left|r-C_\alpha^\ast\right| d P_0+\int_{G^* \backslash \hat{G}}\left|r-C_\alpha^\ast\right| d P_0+C_\alpha^\ast\left\{R_0\left(\phi_\alpha^*\right)-R_0(\hat{\phi}_{\alpha})\right\}.
        \end{aligned}
        \end{equation*}
        The third term could be controlled by Lemma \ref{lemma:type I error deviation}. We focus on bound the first and the second term in the following proof.\\  
        To control the first term we need the following decomposition:
        \begin{equation*}
            \begin{aligned}
                \int_{\hat{G} \backslash G^*}\left|r-C_\alpha^\ast\right| d P_0  
                    =& \int_{(\hat{G} \backslash G^*)\cap \mathcal{C}}\left|r-C_\alpha^\ast\right| d P_0 + \int_{(\hat{G} \backslash G^*) \cap \mathcal{C}^c}\left|r-C_\alpha^\ast \right| d P_0,
            \end{aligned}
        \end{equation*}
        where $\mathcal{C}$ is constructed via Theorem~\ref{thm:deviation of s}.
        Define \[T = \max_{\calX\in\mathcal{C}} | \langle \calX,\calB \rangle - \langle \calX, \hat{\calB\rangle}|,\]
        and 
        \begin{equation*}
            \begin{aligned}
                \Delta R^{+}_{0, \mathcal{C}}=&\left|R_0\left(\phi_\alpha^* \mid \mathcal{C}\right)-R_0\left(\hat{\phi}_{k^*} \mid \mathcal{C}\right)\right| \\
                =&\left|P_0\left(s^*(\calX)>C_\alpha^{**} \mid \calX \in \mathcal{C}\right)-P_0\left(\hat{s}(\calX)>\hat{C}_\alpha \mid \calX \in \mathcal{C}\right)\right|.
            \end{aligned}
        \end{equation*}

        Then by Theorem~\ref{thm:deviation of s}, 
        \begin{equation}
            \label{iqn: upper bound of T}
            T \leq t\sqrt{\frac{d_0 r_0}{n_{\text{min}}}} .
        \end{equation}
        And for $\Delta R^{+}_{0, \mathcal{C}}$:
        \begin{equation*}
            \begin{aligned}
                \Delta R^{+}_{0, \mathcal{C}}
                =& \frac{\left|R_0 (\hat{\phi}_{k^*}) - R_0 (\phi_\alpha^*) - \Pr(\mathcal{C}^c)\left(R_0 (\hat{\phi}_{k^*} \mid \mathcal{C}^c) - R_0 (\phi_\alpha^* \mid \mathcal{C}^c)\right)\right|}{\Pr(\mathcal{C})}\\
                \leq& \frac{|R_0 (\hat{\phi}_{k^*}) - R_0 (\phi_\alpha^*)| + \Pr(\mathcal{C}^c)}{\Pr(\mathcal{C})}
                \stackrel{\eqref{iqn: probabilty of set C}}{\leq} \frac{|R_0 (\hat{\phi}_{k^*}) - R_0 (\phi_\alpha^*)| + 2\exp(-t^2/2)}{1-2\exp(-t^2/2)} \\
                \leq& 2\left( |R_0 (\hat{\phi}_{k^*}) - R_0 (\phi_\alpha^*)| + 2\exp(-t^2/2) \right) .
            \end{aligned}
        \end{equation*}
        Recall lemma \ref{lemma:type I error deviation}, 
        \begin{equation} \label{iqn: conditional excess type I error}
            \Pr\left\{\Delta R^{+}_{0, \mathcal{C}} > 2\left(\xi_{\alpha, \delta_0, {n_0''}}\left(\delta_0^{\prime}\right) + 2\exp(-t^2/2) \right)\right\} \leq \delta_0 +\delta_0' ,
        \end{equation}
        where $\delta_0' \in  (0,1)$ and if $n_0'' \geq \max \left\{\delta_0^{-2}, \delta_0^{\prime-2},4 /\left(\alpha \delta_0\right), \left(\frac{1}{10}M_1(\delta^{*})^{\underline{\gamma}}\right)^{-4}\right\}$, we have \[\xi_{\alpha, \delta_0, n''_0}\left(\delta_0^{\prime}\right) \leq(5 / 2) {n''_0}^{-1 / 4} \leq \frac{1}{4}M_1(\delta^{*})^{\underline{\gamma}}.\] If $t \geq \sqrt{-2\log\left( \frac{1}{8}M_1(\delta^{*})^{\underline{\gamma}} \right)}$, then $2\exp(-t^2/2) \leq \frac{1}{4}M_1(\delta^{*})^{\underline{\gamma}}$. 
        
        From
        \begin{equation*}
            \begin{aligned}
                \Pr \left\{\left(\frac{\Delta R^{+}_{0, \mathcal{C}}}{M_1}\right)^{1 / \underline{\gamma}} > \delta^* \right\}
                \leq \Pr\left\{\Delta R^{+}_{0, \mathcal{C}} > 2\left(\xi_{\alpha, \delta_0, {n_0''}}\left(\delta_0^{\prime}\right) + 2\exp(-t^2/2) \right)\right\} 
                \leq \delta_0 +\delta_0',
            \end{aligned}
        \end{equation*}
        at least with probability $1-\delta_0 -\delta_0'$, we have 
        \begin{equation*} 
            \left(\frac{\Delta R^{+}_{0, \mathcal{C}}}{M_1}\right)^{1 / \underline{\gamma}} \leq \delta^*.
        \end{equation*}
        Next we find the relation between $C_\alpha^{**}$ and $\hat{C_\alpha}$ by applying conditional detection condition (Definition \ref{def:conditional detection condition}).
                \[
                \begin{aligned}
                &P_0\left(s^*(\calX) \geq C_\alpha^{**}+\left(\Delta R^{+}_{0, \mathcal{C}} / M_1\right)^{1 / \underline{\gamma}} \mid \calX \in \mathcal{C}\right) \\
                = & R_0\left(\phi_\alpha^* \mid \mathcal{C}\right)-P_0\left(C_\alpha^{**}<s^*(\calX)<C_\alpha^{**}+\left(\Delta R^{+}_{0, \mathcal{C}} / M_1\right)^{1 / \underline{\gamma}} \mid \calX \in \mathcal{C}\right) \\
                \stackrel{\eqref{iqn: detection condition}}{\leq} & R_0\left(\phi_\alpha^* \mid \mathcal{C}\right)-\Delta R^{+}_{0, \mathcal{C}}  \\
                \leq & R_0\left(\hat{\phi}_{k^*} \mid \mathcal{C}\right)\\
                =&P_0\left(\hat{s}(\calX)>\hat{C}_\alpha \mid \calX \in \mathcal{C}\right) \\
                \leq & P_0\left(s^*(\calX)>\hat{C}_\alpha-T \mid \calX \in \mathcal{C}\right) .
                \end{aligned}
                \]
            Hence,  $C_\alpha^{**}+\left(\Delta R^{+}_{0, \mathcal{C}} / M_1\right)^{1 / \underline{\gamma}}  \geq \hat{C}_\alpha-T$.
            
            Furthermore, 
            \[
            \begin{aligned}
            \left(\hat{G} \backslash G^*\right)
            = &  \left\{s^*>C_\alpha^{**}, \hat{s} \leq \hat{C}_\alpha\right\} \\
            = &  \left\{s^*>C_\alpha^{**}, \hat{s} \leq C_\alpha^{**}+\left(\Delta R^{+}_{0, \mathcal{C}} / M_1\right)^{1 / \underline{\gamma}}+T\right\} \cap\left\{\hat{s} \leq \hat{C}_\alpha\right\} \\
            \subset &   \{C_\alpha^{**}+\left(\Delta R^{+}_{0, \mathcal{C}} / M_1\right)^{1 / \underline{\gamma}}+2 T \geq s^* \geq C_\alpha^{**}, 
             \hat{s} \leq C_\alpha^{**}+\left(\Delta R^{+}_{0, \mathcal{C}} / M_1\right)^{1 / \underline{\gamma}}+T\} \\
             &\cap\left\{\hat{s} \leq \hat{C}_\alpha\right\} \\
            \subset &  \left\{C_\alpha^{**}+\left(\Delta R^{+}_{0, \mathcal{C}} / M_1\right)^{1 / \underline{\gamma}}+2 T \geq s^* \geq C_\alpha^{**}\right\}.
            \end{aligned}
            \]
            Denote $\mathcal{C}_1=\left\{\calX: C_\alpha^{**}+\left(\Delta R^{+}_{0, \mathcal{C}} / M_1\right)^{1 / \underline{\gamma}}+2 T \geq s^*(\calX) \geq C_\alpha^{**}\right\}$.
            Then the integration on $(\hat{G} \backslash G^*)\cap \mathcal{C}$ could be bounded via the Mean Value Theorem (recalling that
            $r(\calX) = f_1(\calX)/f_0(\calX)=\exp\left( s^*(\calX)-\langle \cM, \; \calB \rangle \right)$ and $ C_\alpha= \exp\left( C_\alpha^{\ast\ast} - \langle \cM, \; \calB \rangle \right)$):
            $$
            \left|r(\calX)-C_\alpha^\ast\right|=e^{-\langle \cM, \; \calB \rangle}\left|e^{s^*(\calX)}-e^{C_\alpha^{**}}\right|=e^{-\langle \cM, \; \calB \rangle} \cdot e^{z^{\prime}}\left|s^*(\calX)-C_\alpha^{**}\right|,
            $$
            where $z^{\prime}$ is some quantity between $s^*(\calX)$ and $C_\alpha^{**}$.  Restricting to $\mathcal{C} \cap \mathcal{C}_1$, we have
            $$
            C_\alpha^{**} \leq z^{\prime} \leq C_\alpha^{**}+\left(\Delta R^{+}_{0, \mathcal{C}} / M_1\right)^{1 / \underline{\gamma}}+2 T .
            $$
            And
            $$
            \begin{aligned}
            &\int_{(\hat{G} \backslash G^*)\cap \mathcal{C}}\left|r-C_\alpha^\ast\right| d P_0  \\
             \leq& \int_{\mathcal{C} \cap \mathcal{C}_1}\left|r-C_\alpha^\ast\right| d P_0 \\
             =& \int_{\mathcal{C} \cap \mathcal{C}_1} \exp \left\{z^{\prime}-\langle \cM, \; \calB \rangle\right\}\left|s^*(\calX)-C_\alpha^{**}\right| d P_0 \\
            \leq& \int_{\mathcal{C} \cap \mathcal{C}_1} \exp \left\{C_\alpha^{**}+\left(\Delta R^{+}_{0, \mathcal{C}} / M_1\right)^{1 / \underline{\gamma}}+2 T-\langle \cM, \; \calB \rangle\right\}\left|s^*(\calX)-C_\alpha^{**}\right| d P_0 \\
            \leq & c^{\prime} \int_{\mathcal{C}\cap \mathcal{C}_1}\left|s^*(\calX)-C_\alpha^{**}\right| d P_0 \\
            \leq & c^{\prime}\left(\left(\Delta R^{+}_{0, \mathcal{C}} / M_1\right)^{1 / \underline{\gamma}}+2 T\right) P_0\left(\mathcal{C} \cap \mathcal{C}_1\right).
            \end{aligned}
            $$
            On the other hand, 
            $$
            \begin{aligned}
            \int_{(\hat{G} \backslash G^*)\cap \mathcal{C}^c}\left|r-C_\alpha^\ast\right| d P_0 
            \leq& \int_{\mathcal{C}^c}\left|r-C_\alpha^\ast\right| d P_0 
            \leq \int_{\mathcal{C}^c} r d P_0+C_\alpha^\ast \int_{\mathcal{C}^c} d P_0
            \\=&P_1\left(\mathcal{C}^c\right)+C_\alpha^\ast P_0\left(\mathcal{C}^c\right) 
            \leq 2\left(1+C_\alpha^\ast\right) \exp \left\{-t^2/2\right\}. 
            \end{aligned}
            $$
            Together with $P_0\left(\mathcal{C} \cap \mathcal{C}_1\right) \stackrel{\eqref{iqn: margin assumption}}{\leq} M_0 \left(\left(\Delta R^{+}_{0, \mathcal{C}} / M_1\right)^{1 / \underline{\gamma}}+2 T\right)^{\bar{\gamma}}$,
            \begin{equation}\label{iqn: the first term}
                \begin{aligned}
                    \int_{\left(\hat{G} \backslash G^*\right)}\left|r-C_\alpha^\ast\right| d P_0 
                    \leq & c^{\prime} M_0\left(\left(\Delta R^{+}_{0, \mathcal{C}} / M_1\right)^{1 / \underline{\gamma}}+2 T\right)^{1+\bar{\gamma}}+2\left(1+C_\alpha^\ast\right) \exp \left\{-t^2/2\right\}.
                \end{aligned}
            \end{equation}

        To derive an upper bounnd of the second term, we consider the same decomposition:
        \begin{equation*}
            \begin{aligned}
                \int_{G^* \backslash \hat{G}}\left|r-C_\alpha^\ast\right| d P_0  
                    =& \int_{(G^* \backslash \hat{G})\cap \mathcal{C}}\left|r-C_\alpha^\ast\right| d P_0 + \int_{(G^* \backslash \hat{G}) \cap \mathcal{C}^c}\left|r-C_\alpha^\ast\right| d P_0 .
            \end{aligned}
        \end{equation*}
        Notice that on $G^* \backslash \hat{G}$ we have $s^*(\calX)<C_\alpha^{**}$, hence we need to derive an lower bound of $s^*(\calX)$ (conditional on $G^* \backslash \hat{G}$), in order to apply the Mean Value Theorem. Define \[ \Delta R_{0, \mathcal{C}}=R_0\left(\phi_\alpha^* \mid \mathcal{C}\right)-R_0\left(\hat{\phi}_{k^*} \mid \mathcal{C}\right)=P_0\left(s^*(\calX)>C_\alpha^{**} \mid \calX \in \mathcal{C}\right)-P_0\left(\hat{s}(\calX)>\hat{C}_\alpha \mid \calX \in \mathcal{C}\right).\]
        Then we have $\Delta R^{+}_{0, \mathcal{C}} = |\Delta R_{0, \mathcal{C}}|$. Next we discuss the cases $\Delta R_{0, \mathcal{C}}\geq0$ and $\Delta R_{0, \mathcal{C}}<0$ seperately:\\
        If $\Delta R_{0, \mathcal{C}}\geq0$, by applying conditional margin assumption (Definition \ref{def:conditional margin assumption}):
            $$
            \begin{aligned}
            & P_0\left(s^*(\calX) \geq C_\alpha^{**}+\left(\Delta R_{0, \mathcal{C}} / M_0\right)^{1 / \bar{\gamma}} \mid \calX \in \mathcal{C}\right) \\
            = & R_0\left(\phi_\alpha^* \mid \mathcal{C}\right)-P_0\left(C_\alpha^{**}<s^*(\calX)<C_\alpha^{**}+\left(\Delta R_{0, \mathcal{C}} / M_0\right)^{1 / \bar{\gamma}} \mid \calX \in \mathcal{C}\right) \\
            \stackrel{\eqref{iqn: margin assumption}}{\geq} & R_0\left(\phi_\alpha^* \mid \mathcal{C}\right)-\Delta R_{0, \mathcal{C}}  \\
            =&P_0\left(\hat{s}(\calX)>\hat{C}_\alpha \mid \calX \in \mathcal{C}\right) \\
            \geq & P_0\left(s^*(\calX)>\hat{C}_\alpha+T \mid \calX \in \mathcal{C}\right) .
            \end{aligned}
            $$
             Hence,  $C_\alpha^{**}+\left(\Delta R^{+}_{0, \mathcal{C}} / M_0\right)^{1 / \bar{\gamma}}  \leq \hat{C}_\alpha+T$.
             
        If $\Delta R_{0, \mathcal{C}}<0$, by applying conditional detection condition (Definition \ref{def:conditional detection condition}):
            $$
            \begin{aligned}
            & P_0\left(s^*(\calX) \geq C_\alpha^{**}-\left(-\Delta R_{0, \mathcal{C}} / M_1\right)^{1 / \underline{\gamma}} \mid \calX \in \mathcal{C}\right) \\
            = & R_0\left(\phi_\alpha^* \mid \mathcal{C}\right)+P_0\left(C_\alpha^{**}-\left(-\Delta R_{0, \mathcal{C}} / M_1\right)^{1 / \underline{\gamma}}<s^*(\calX)<C_\alpha^{**} \mid \calX \in \mathcal{C}\right) \\
            \stackrel{\eqref{iqn: detection condition}}{\geq} & R_0\left(\phi_\alpha^* \mid \mathcal{C}\right)-\Delta R_{0, \mathcal{C}}  \\
            =&P_0\left(\hat{s}(\calX)>\hat{C}_\alpha \mid \calX \in \mathcal{C}\right) \\
            \geq & P_0\left(s^*(\calX)>\hat{C}_\alpha+T \mid \calX \in \mathcal{C}\right) .
            \end{aligned}
            $$
            Hence,  $C_\alpha^{**}-\left(\Delta R^{+}_{0, \mathcal{C}} / M_1\right)^{1 / \underline{\gamma}}  \leq \hat{C}_\alpha+T$.\\
            In general, 
            \[\hat{C}_\alpha \geq C_\alpha^{**}-\left(\Delta R^{+}_{0, \mathcal{C}} / M_1\right)^{1 / \underline{\gamma}}\wedge\left(\Delta R^{+}_{0, \mathcal{C}} / M_0\right)^{1 / \bar{\gamma}}  -T .\]
            Furthermore, 
            $$
            \begin{aligned}
            \left(G^* \backslash \hat{G}\right)
            = & \left\{s^*\leq C_\alpha^{**},\; \hat{s} > \hat{C}_\alpha\right\} \\
            = &  \left\{s^*\leq C_\alpha^{**},\; \hat{s} > C_\alpha^{**}-\left(\Delta R^{+}_{0, \mathcal{C}} / M_1\right)^{1 / \underline{\gamma}}\wedge\left(\Delta R^{+}_{0, \mathcal{C}} / M_0\right)^{1 / \bar{\gamma}}  -T \right\} \\
            &\cap\left\{\hat{s} > \hat{C}_\alpha\right\} \\
            \subset &   \{C_\alpha^{**}-\left(\Delta R^{+}_{0, \mathcal{C}} / M_1\right)^{1 / \underline{\gamma}}\wedge\left(\Delta R^{+}_{0, \mathcal{C}} / M_0\right)^{1 / \bar{\gamma}}  -2T \leq s^* \leq C_\alpha^{**}\}. 
            \end{aligned}
            $$
            Denote $\mathcal{C}_2=\left\{C_\alpha^{**}-\left(\Delta R^{+}_{0, \mathcal{C}} / M_1\right)^{1 / \underline{\gamma}}\wedge\left(\Delta R^{+}_{0, \mathcal{C}} / M_0\right)^{1 / \bar{\gamma}}  -2T \leq s^* \leq C_\alpha^{**}\right\}$.
            Then the integration on $\left(G^* \backslash \hat{G}\right)\cap \mathcal{C}$ could be bounded via the Mean Value Theorem (recalling that
            $r(\calX)=\exp\left( s^*(\calX)-\langle \cM, \; \calB \rangle \right)$ and $ C_\alpha^\ast= \exp\left( C_\alpha^{\ast\ast} - \langle \cM, \; \calB \rangle \right)$):
            $$
            \left|r(\calX)-C_\alpha^\ast\right|=e^{-\langle \cM, \; \calB \rangle}\left|e^{s^*(\calX)}-e^{C_\alpha^{**}}\right|=e^{-\langle \cM, \; \calB \rangle} \cdot e^{z^{\prime}}\left|s^*(\calX)-C_\alpha^{**}\right|,
            $$
            where $z^{\prime}$ is some quantity between $s^*(\calX)$ and $C_\alpha^{**}$.  Restricting to $\mathcal{C} \cap \mathcal{C}_2$, we have
            $$
            C_\alpha^{**}-\left(\Delta R^{+}_{0, \mathcal{C}} / M_1\right)^{1 / \underline{\gamma}}\wedge\left(\Delta R^{+}_{0, \mathcal{C}} / M_0\right)^{1 / \bar{\gamma}}  -2T \leq z^{\prime} \leq C_\alpha^{**}.
            $$
            And
            $$
            \begin{aligned}
            \int_{\left(G^* \backslash \hat{G}\right)\cap \mathcal{C}}\left|r-C_\alpha^\ast\right| d P_0
            \leq& \int_{\mathcal{C} \cap \mathcal{C}_2}\left|r-C_\alpha^\ast\right| d P_0 \\
            =& \int_{\mathcal{C} \cap \mathcal{C}_2} \exp \left\{z^{\prime}-\langle \cM, \; \calB \rangle\right\}\left|s^*(\calX)-C_\alpha^{**}\right| d P_0 \\
            \leq& \int_{\mathcal{C} \cap \mathcal{C}_2} \exp \left\{C_\alpha^{**}-\langle \cM, \; \calB \rangle\right\}\left|s^*(\calX)-C_\alpha^{**}\right| d P_0 \\
            \leq & c^{\prime\prime} \int_{\mathcal{C}\cap \mathcal{C}_2}\left|s^*(\calX)-C_\alpha^{**}\right| d P_0 \\
            \leq & c^{\prime\prime}\left(\left(\Delta R^{+}_{0, \mathcal{C}} / M_1\right)^{1 / \underline{\gamma}}\wedge\left(\Delta R^{+}_{0, \mathcal{C}} / M_0\right)^{1 / \bar{\gamma}} +2T\right) P_0\left(\mathcal{C} \cap \mathcal{C}_2\right).
            \end{aligned}
            $$
            On the other hand, 
            $$
            \begin{aligned}
            \int_{\left(G^* \backslash \hat{G}\right)\cap \mathcal{C}^c}\left|r-C_\alpha^\ast\right| d P_0 
            \leq& \int_{\mathcal{C}^c}\left|r-C_\alpha^\ast\right| d P_0 
            \leq \int_{\mathcal{C}^c} r d P_0+C_\alpha^\ast \int_{\mathcal{C}^c} d P_0
            \\=& P_1\left(\mathcal{C}^c\right)+C_\alpha^\ast P_0\left(\mathcal{C}^c\right)
            \leq 2\left(1+C_\alpha^\ast\right) \exp \left\{-t^2/2\right\}.
            \end{aligned}
            $$
            Together with $P_0\left(\mathcal{C} \cap \mathcal{C}_2\right)  \stackrel{\eqref{iqn: margin assumption}}{\leq} M_0 \left(\left(\Delta R^{+}_{0, \mathcal{C}} / M_1\right)^{1 / \underline{\gamma}}\wedge\left(\Delta R^{+}_{0, \mathcal{C}} / M_0\right)^{1 / \bar{\gamma}} +2T\right)^{\bar{\gamma}}$,
            \begin{equation}\label{iqn: the second term}
                \begin{aligned}
                    &\int_{\left(G^* \backslash \hat{G}\right)}\left|r-C_\alpha^\ast\right| d P_0 \\
                    \leq & c^{\prime\prime} M_0\left(\left(\Delta R^{+}_{0, \mathcal{C}} / M_1\right)^{1 / \underline{\gamma}}\wedge\left(\Delta R^{+}_{0, \mathcal{C}} / M_0\right)^{1 / \bar{\gamma}} +2T\right)^{1+\bar{\gamma}}+2\left(1+C_\alpha^\ast \right) \exp \left\{-t^2/2\right\}.
                \end{aligned}
            \end{equation}
        
            Combine \eqref{eqn:excess-typeii-error}, Lemma \ref{lemma:type I error deviation}, \eqref{iqn: the first term}, \eqref{iqn: the second term}, \eqref{iqn: upper bound of T}, \eqref{iqn: conditional excess type I error}, one could conclude:
            Suppose Assumption~\ref{aspt:bounded norm and snr} as well as the assumptions in Proposition~\ref{theorem: tucker} hold, and further assume $t \geq \sqrt{-2\log\left( \frac{1}{8}M_1(\delta^{*})^{\underline{\gamma}} \right)}$, $n_0'' \geq \max \left\{\delta_0^{-2}, \delta_0^{\prime}-2,4 /\left(\alpha \delta_0\right), \left(\frac{1}{10}M_1(\delta^{*})^{\underline{\gamma}}\right)^{-4}\right\}$, where $\delta_0' \in  (0,1)$, 
            {\small \begin{equation*}
                \begin{aligned}
                 t\asymp& (n_0\wedge n_1)^a \\
                 \geq& \max\left\{\Phi^{-1}(1-\alpha), C_0^M\sqrt{1+3\frac{C_B}{C_L}}\Phi^{-1}(\frac{3}{4})+\frac{C_0^{3M/2}C_B}{C_L}\left(\delta^*+\sqrt{C_L}\Phi^{-1}(1-\alpha)+C_L/2\right)\right\},
                \end{aligned}
            \end{equation*}}
            and $d_0 \asymp (n_0\wedge n_1)^b$, then with probability at least $1-\delta_0-\delta_0^{\prime} - C\exp(-cd_0)$
            {\small\begin{equation*}
            \begin{aligned}
            & {\Pr}_1(\hat{G})-{\Pr}_1\left(G^*\right) \\
            = &\int_{\hat{G} \backslash G^*}\left|r-C_\alpha^\ast\right| d P_0+\int_{G^* \backslash \hat{G}}\left|r-C_\alpha^\ast\right| d P_0+C_\alpha^\ast\left\{R_0\left(\phi_\alpha^*\right)-R_0(\hat{\phi}_{k^*})\right\}\\
            \leq & c^{\prime} M_0\left(\left(\Delta R^{+}_{0, \mathcal{C}} / M_1\right)^{1 / \underline{\gamma}}+2 T\right)^{1+\bar{\gamma}}+2\left(1+C_\alpha^\ast\right) \exp \left\{-t^2/2\right\} \\
            &\quad +c^{\prime\prime} M_0\left(\left(\Delta R^{+}_{0, \mathcal{C}} / M_1\right)^{1 / \underline{\gamma}}\wedge\left(\Delta R^{+}_{0, \mathcal{C}} / M_0\right)^{1 / \bar{\gamma}} +2T\right)^{1+\bar{\gamma}}\\
            &\quad +2\left(1+C_\alpha^\ast\right) \exp \left\{-t^2/2\right\} 
            + C_\alpha^\ast \xi_{\alpha, \delta_0, {n_0''}}\left(\delta_0^{\prime}\right)\\
            \leq& C_1 \left(\Delta R^{+}_{0,\mathcal{C}}\right)^{(1+\bar{\gamma})/\underline{\gamma}}
            +C_2 \left(\Delta R^{+}_{0,\mathcal{C}}\right)^{(1+\bar{\gamma})/\bar{\gamma}}
            +C_3 T^{1+\bar{\gamma}}
            +C_4 \exp \left\{-t^2/2\right\}
            +C_\alpha^\ast \xi_{\alpha, \delta_0, {n_0''}}\left(\delta_0^{\prime}\right)\\
            \leq& C_1'\left(\xi_{\alpha, \delta_0, {n_0''}}\left(\delta_0^{\prime}\right)\right)^{(1+\bar{\gamma})/\underline{\gamma}} 
            +C_2'\left(\xi_{\alpha, \delta_0, {n_0''}}\left(\delta_0^{\prime}\right)\right)^{(1+\bar{\gamma})/\bar{\gamma}}
            +C_\alpha^\ast \xi_{\alpha, \delta_0, {n_0''}}\left(\delta_0^{\prime}\right)\\
            &\quad + C_3'\left(\exp(-t^2/2) \right)^{(1+\bar{\gamma})/\underline{\gamma}} 
            + C_4'\left(\exp(-t^2/2) \right)^{(1+\bar{\gamma})/\bar{\gamma}}
            + C_5' \exp \left\{-t^2/2\right\} \\
            &\quad + C_6' \left( t\sqrt{\frac{d_0 r_0}{n_{\text{min}}}} \right)^{1+\bar{\gamma}}\\
            \leq& C_1'' ({n_0''})^{-\min\left\{\frac{1+\bar{\gamma}}{4\underline{\gamma}}, \frac{1+\bar{\gamma}}{4\bar{\gamma}}, \frac{1}{4} \right\}}
            + C_2''e^{-\frac{t^2}{2}\min\left\{\frac{1+\bar{\gamma}}{\underline{\gamma}}, \frac{1+\bar{\gamma}}{\bar{\gamma}}, 1\right\}}
            +C_3'' \left( t\sqrt{\frac{d_0 r_0}{n_{\text{min}}}} \right)^{1+\bar{\gamma}},
            \end{aligned}
            \end{equation*}}
            where $C_1, C_2, C_3, C_4, C_1', C_2', C_3', C_4', C_5', C_6', C_1'', C_2'', C_3''$ are constants.
    \end{proof}

\end{document}